\documentclass[twocolumn]{aastex631}
\usepackage{xspace}
\usepackage{xcolor}
\definecolor{dark green}{rgb}{0.0,0.5,0.0}
\usepackage{placeins} % Allows use of \FloatBarrier to control float placement
\usepackage{amssymb} % For \simeq (≃) symbol
\usepackage{amsmath}
\usepackage[utf8]{inputenc} % Allow Unicode characters
\usepackage{ltablex}
\usepackage{array}
\usepackage{booktabs}
\usepackage{ragged2e}
\usepackage{float}

\DeclareRobustCommand{\ion}[2]{%
\relax\ifmmode
\ifx\testbx\f@series
{\mathbf{#1\,\mathsc{#2}}}\else
{\mathrm{#1\,\mathsc{#2}}}\fi
\else\textup{#1\,{\mdseries\textsc{#2}}}%
\fi}

\newcommand{\Ms}{\ensuremath{M_\odot}}

\newcommand{\ec}{$\eta$\,Car\xspace}

\DeclareRobustCommand{\ion}[2]{%
\relax\ifmmode
\ifx\testbx\f@series
{\mathbf{#1\,\mathsc{#2}}}\else
{\mathrm{#1\,\mathsc{#2}}}\fi
\else\textup{#1\,{\mdseries\textsc{#2}}}%
\fi}

\begin{document}
\title[rolof]{
Eta Carinae's historical light curve: evidence for cyclic Roche lobe overflow from the primary star
} 

\correspondingauthor{Damineli, Augusto}
\email{augusto.damineli@gmail.com},
\author[0000-0002-7978-2994]{Damineli,~Augusto}
\affiliation{Instituto de Astr., Geof\'isica e Ci\^encias Atmosf. da USP, Rua do Mat\~ao 1226, Cidade Universit\'aria, S\~ao Paulo, Brasil}

\author[0000-0002-3817-6402]{Almeida,~L. A.}
\affiliation{Escola de Ciências e Tecnologia, Universidade Federal do Rio Grande do Norte, Campus Universitário, 59092-970, Natal, RN, Brazil}

\author[0000-0002-0386-2306]{Jablonski, ~Francisco ~J.}
\affiliation{Instituto Nac. de Pesq. Espaciais/MCTI Avenida dos Astronautas 1758, S\~ao Jos\'e dos Campos, SP, 12227-010, Brazil}

\author[0000-0002-9262-4456]{Fern\'andez-Laj\'us, ~Eduardo}
\affiliation{Instituto de Astrofísica de La~Plata (CCT La~Plata - CONICET/UNLP), Argentina}

\author[0000-0002-0284-0578]{Navarete, ~Felipe}
\affiliation{Laboratório Nacional de Astrofísica, Rua dos Estados Unidos 154, 37504-364, Itajubá, MG, Brazil}

\author[0000-0002-0284-0578]{Martioli,~Eder}
\affiliation{Laboratório Nacional de Astrofísica, Rua dos Estados Unidos 154, 37504-364, Itajubá, MG, Brazil}

\author[0000-0001-9754-2233]{Weigelt, ~Gerd}
\affiliation{Max Planck Institute for Radio Astronomy, Auf dem H\"{u}gel 69, D-53121 Bonn, Germany}

\author[0009-0005-0701-9403]{Capobiango, ~Rodrigo}
\affiliation{Instituto de Astr., Geof\'isica e Ci\^encias Atmosf. da USP, Rua do Mat\~ao 1226, Cidade Universit\'aria, S\~ao Paulo, Brazil}

\begin{abstract}
The large amount of ground-based photometric measurements of $\eta$ Carinae obtained since 1940 have remained problematic for quantitative modeling due to the blending of flux from the stellar core and the surrounding circumstellar nebula. In the era of the Hubble Space Telescope, spatially resolved imaging \& spectrophotometry have enabled disentanglement of these components, allowing recovery of the stellar core $V$-band brightness from ground-based observations. We isolate the $V$-band fluxes of the stellar core and nebula using 1999--2020 HST (ACS and STIS) observations, and use these to calibrate coeval ground-based photometry. The main finding is an orbital light curve with an amplitude $\Delta m \approx \pm 0.2$\,mag, many times higher than that modeled by ellipsoidal deformation of the primary. The observations suggest that Roche lobe overflow starts  at $- 75$ days before periastron in coincidence with the start of rising in the orbital light curve, and remains for 150 days. An expanding (and afterwards dissipating) gas cloud reflecting the light from the primary would explain the observed large amplitude of the orbital light curve. A sharp periodic photometric peak occurs at {$\sim -18$} days from the periastron.  It is followed by a broad minimum around the superior conjunction of the secondary (T$_0+5.2$ days), which we interpret as a partial eclipse of the ejected material, in coincidence with the \emph{shallow minimum} in X-rays, which also has been attributed to an eclipse.
\end{abstract}


\section{Introduction}
\label{introduction}

The photometric evolution of eta~Carinae {(\ec)} after 1940 has not yet received the attention it deserves. In particular, no comprehensive review has been made in light of our current understanding of the system, and the major episodes of variability have not been clearly re-examined in connection with the correlated spectroscopic events. This task cannot be fully addressed within the scope of this paper. Instead, we focus on the main structural aspects of the system and, through our data analysis, propose a quantitative interpretation of the photometric events that occurred during the 1963–2026 interval.

The central engine of \ec consists of a highly eccentric binary system of long-period \citep{damineli96}, with $e \sim 0.9$ and $P = 5.54$\,yr \citep{damineli97, damineli00, grant20}. Orbital modulation is observed across essentially all wavelength ranges \citep{damineli08a}: in X-rays, explained by the strong collision between the stellar winds of the companion stars \citep{corcoran09, pittard02, hamaguchi14, espinoza22}; in the near infrared, powered by free-free emission of the wind \citep{whitelock04, feast01, mehner14}; and in the ultraviolet, produced by gas illuminated by the secondary star \citep{Gull22}. Consistent interpretations suggest that the present binary system is the remnant of a former triple system in which two stars merged during the 1840s eruption \citep{smith18c,hirai21}, ejecting a large amount of mass and producing the Homunculus Nebula, including a dusty torus of $\sim 45$\,\Ms \citep{morris17} and a polar mass of $\sim 2.5$\,\Ms \citep{smith03c}. 

\begin{figure*}[!ht]
\centering
{\includegraphics[width=\linewidth,angle=0,viewport=0bp 0bp 1400bp 820bp, clip]{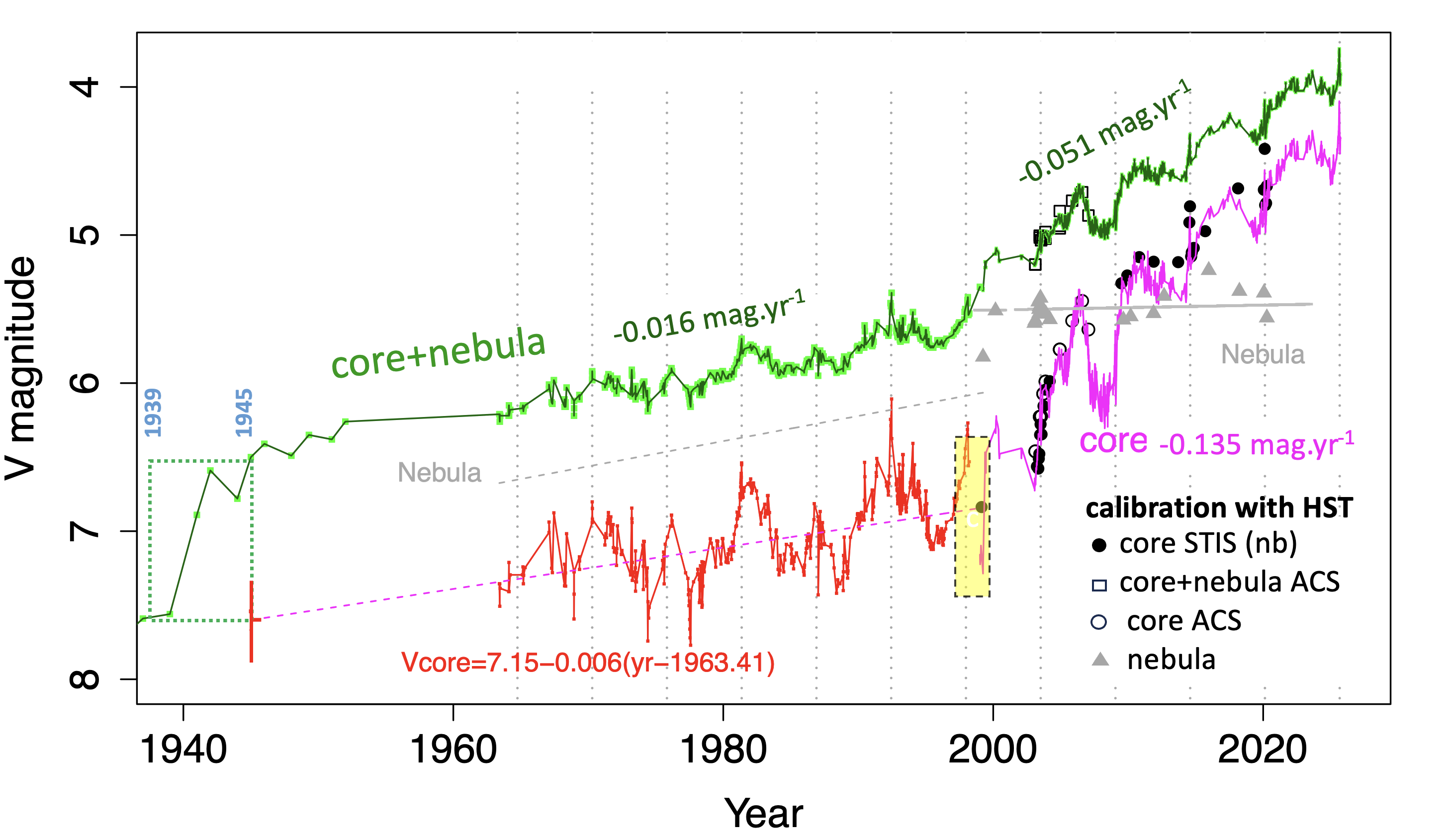}}
\caption{ $V$-band photometry of \ec in the time-frame (1963--2026). {\it Upper green curve:} $V$-mag for the whole object.  {\it Lower magenta and red plots:} calibrated $V$-mag for the central stellar core (following \citet[][see their Sect.~3.2]{damineli19}). {\it Polygons:} HST(STIS+ACS); {\it gray lines:} nebular $V$-mag - see text. The vertical gray dashed lines mark the times of periastron passage. The magnitude of the stellar core in 1945 - \cite{oconnell56} - is quite uncertain ($7.6\pm0.3$\,mag), but its impact is lessened when applied to the gradient for the subsequent 60 years. The analysis is split into two epochs: before and after 1999, when there was a conspicuous brightness jump.} 
\label{Fig:Vall}
\end{figure*}

The time window (Fig.\ref{Fig:Vall}) analyzed in this work begins with a sudden brightening of 1.1\,mag in the optical in 1940 within an interval of only seven months. Short-exposure photographic plates indicate that the central star(s) itself did not significantly increase in brightness during that event \citep{oconnell56, thackeray53}. From that point onward, the object as a whole (stellar core + Homunculus) underwent a slow secular brightening, with occasional oscillations and an almost constant {($B-V$)} color index. Spectroscopic events characterized by the disappearance of high-excitation emission lines {([\ion{Fe}{iii}], [\ion{Ne}{iii}], [\ion{Ar}{iii}] and [\ion{S}{iii}])} were recorded in 1948 by \citet{gaviola53} and in several subsequent years. These low-excitation events were  later demonstrated to be periodic (5.54\,yr) by \citet{damineli96}, in correspondence to sharp photometric peaks in {$JHKL$} and to the simultaneous disappearance of the {\ion{He}{i} ($\lambda\,10830$\,{\AA})} line. modeling of the Keplerian velocities derived from the Pa\,$\gamma$ and Pa\,$\delta$ lines revealed a system composed of two massive stars in a highly eccentric orbit, suggesting a colliding-wind binary with X-ray-emitting plasma at temperatures of tens of millions of Kelvin \citep{damineli97}. X-ray monitoring subsequently revealed a remarkably stable light curve, with a pronounced minimum at the periastron \citep{hamaguchi14,espinoza22}.

Mid-infrared photometry \citep{mehner19} also revealed a nearly flat light curve, with very little long-term variability. The emission in this spectral range arises from the reprocessing of ultraviolet radiation from the binary system by dust in the Homunculus. {Therefore, both light curves} indicate that the central source itself is not subject to significant long-term variability.

This behavior is clearly in contrast to the strong brightening observed at UV–optical–NIR wavelengths (a factor of $\sim 10$ over 30 \, year). At the same time, substantial spectroscopic evolution took place, which some authors interpreted as evidence for intrinsic evolution of the central star(s). However, this evolution proved to be largely restricted to the central source, whose apparent brightness increased much faster than that of the Homunculus \citep{martin04}.

\citet{hillier92} had already pointed out that the equivalent width (EW) of the {H$\alpha$} line measured directly at the central source was much larger than the EW measured in the reflected spectrum of the Homunculus Nebula. This suggested the presence of a dusty coronagraph producing additional extinction toward the primary's stellar disk. Later measurements  \citep{mehner12,damineli21} showed that the EW of the H$\alpha$ line measured directly on the star gradually decreased until it became comparable to that observed in the reflected nebular light, indicating that the occulter had nearly completely dissipated \citep{damineli19,gull23}. Furthermore, unusually bright objects at separations of 0\farcs1--0\farcs3 also suggested a coronagraph-like structure \citep{weigelt95}.

The secular brightening observed during the 1999--2026 interval is therefore restricted to a decrease in circumstellar extinction at $d \approx 900~to~700$\,AU \citep{damineli24} along our particular line of sight, unlike in the 1936–1945 interval, when the extinction decreased at much larger distances from the stellar core and apparently in all directions.

There are low-amplitude pulsations with period $P = 58.8$\,d, induced by tidal interaction with the secondary star discussed by \citet{richardson18}, but these are outside the scope of this work.

This work is organized as follows: In Section~\ref{data} we show the origin of data used in this work; in Section~\ref{results} we describe: the photometry of the whole object (stellar core + Homunculus Nebula) in the time-frame 1963--1999, the extraction of the stellar flux in 1999--2026, the orbital light curve of the stellar core in the last 4 orbital cycles, the pre-periastron peak, our modeling of the orbital modulation \textcolor{blue}{suggesting Roche lobe overflow, plus} the determination of the size of evaporating dust in 1999--2026; in Section~\ref{discussions} we present our remarks and conclusions.

\section{Data: observations and reduction}
\label{data}

We started by collecting all published ground-based $V$-band magnitudes since 1963 for the whole object (stellar core + Homunculus nebula). In addition, we report unpublished photometry. They were collected at the Pico dos Dias-Br observatory ($V-band$), La Plata-Ar ($BVRI$) and CASLEO-Ar ($UBVRI$).   The tables will appear in CDS. These were supplemented by magnitudes published in AAVSO database\footnote{{https://aavso.org}}.

All ground-base data are shown in Fig.~\ref{Fig:Vall} as green tiny dots+solid line. The magnitudes in the period 1939--1963 are from \citet{frew04} for which we adopted generous uncertainties. We made small adjustments to bring the $V$ magnitudes to the La Plata system \citep{lajus09b}, adopting the zero-point $V=8.17$ for the comparison star {HD \, 303308} \citep{damineli19} such that the magnitudes obtained with the HST/ACS/HRC images in the F550M filter transform to the magnitudes of the whole object in the La Plata system. We applied the same {photometric} zero-point to the stellar object and derived the flux for the nebular contribution of coeval ground-based images. These three components are shown as open polygons in Fig.~\ref{Fig:Vall} (years 1999--2006).

In addition, we performed narrow-band synthetic photometry on the HST/STIS stellar spectra (filled polygons in Fig.~\ref{Fig:Vall}) to obtain the corresponding stellar magnitudes (filled circles). We used these stellar magnitudes to subtract the stellar flux from the ground-based magnitudes to derive the nebular brightness (gray triangles in Fig.\ref{Fig:Vall}). The magnitudes of the nebula are amazingly constant at $V\approx 5.5$\,mag -- this is because the core brightening of the last three decades was restricted to our line-of-sight and not \emph{seen} by other directions in the nebula. The fit – almost a constant – is represented by a straight, slightly inclined gray line in Fig. 1.

\subsection{Disentangling the stellar core magnitude in ground-based observations}

The magnitude of the stellar core can be recovered from seeing-limited ground-based observations when an independent estimate of the nebular brightness is available. Although this approach may appear indirect, it is more robust than methods based on aperture photometry because the nebular emission is largely insensitive to the short-term photometric variations of the stellar core. As a result, the nebular flux provides a stable reference that can be used to isolate the contribution of the central source \citep{damineli19}.

Under the assumption that the observed flux is the sum of the stellar core and nebular components, the stellar core flux is obtained by subtracting the nebular flux from the total flux of the system:
\begin{equation}
%\mathrm{m_{core} = -2.5 \log_{10}\left(10^{-0.4\mathrm{m_{whole}}} - 10^{-0.4\, m_{neb}}\right)
V_{core} = -2.5 \log_{10} \left(10^{-0.4 V_{whole}} - 10^{-0.4\,V_{neb}} \right)
\end{equation}

\section{Results}
\label{results}

\subsection{The light curve of the stellar core in the time-frame 1963--1999}

Although the light curve of \ec  (core + nebula) from 1940 onward is documented with frequent measurements of good photometric quality, it has not received adequate attention or numerical modeling. After the Minor Eruption (1865), the object remained approximately constant in brightness until 1939.

In 1940, the brightness increased sharply by $\Delta V \approx-1.1$\,mag in seven years. The photographic magnitudes of the stellar core reported by \citet{vandenBos1938} were $\approx 7.9$\,mag in 1932--1936 \citep{vandenBos1938}.
 The stellar core after 1939 brightened much less than the nebula  \citep{oconnell56, thackeray53}.  In this work we adopt the zero-point shift from photographic to $m_V$ reported by \citet{frew04} - their table 4 -  to which we associated a large uncertainty: $\mathrm{m}_V = 7.6 \pm 0.3$\,mag in 1946. The magnitude of the stellar core in 1999.45 was measured in the HST spectrum as $V_{core}=6.9\pm0.07$ \citep{damineli19}. Using these two points, we derived the average stellar core brightening in the window 1963-1999 as:
%$\mathrm{<V_{core}>} = 7.15 - 0.006\,\mathrm{(yr-1963.41)}$.    (eq.~2)\\
\begin{equation}
\langle V_{core}\rangle = 7.15 - 0.006(\mathrm{yr}-1963.41).
\end{equation}
 
 Using the average $\langle V_{core} \rangle$ in the time window 1963--1999, $V_{nebula}$ at every date can be derived as:
%$\langle V_{\mathrm{nebula}} \rangle = -2.5 \log_{10}\left(10^{-0.4 \langle V_{\mathrm{whole}} \rangle} - 10^{-0.4 \langle V_{\mathrm{core}} \rangle}\right)$   (eq.~3)\\
\begin{equation}
    \langle V_{neb} \rangle = -2.5 \log_{10}\left(10^{-0.4 \langle V_{whole} \rangle} - 10^{-0.4 \langle V_{core} \rangle}\right)
\end{equation}
The nebular contribution was then subtracted from the whole object brightness (green line in Fig.~\ref{Fig:Vall}) to get the magnitudes of the stellar core at every date -- the red line. Of course, this simplified procedure ignores possible local fluctuations of the nebular brightness, which we believe to be minor when looking to the subsequent time window 1999--2026 -- see the next subsection.
 
The 1940 brightness jump was probably an episode of extinction decrease of the nebula in all directions. In particular, this produced an increase in the ionization of Weigelt globules \citep{weigelt86,hofmann88}, which exhibit high-ionization narrow-line emission features.
The temporary disappearance of these high-excitation lines led to the discovery that it was periodic \citep{damineli96}, which was later associated with the eccentric binary system by \citet{damineli97}. From 1963 to 1999, the whole object experienced a slow increase in brightness, with $\delta V=-0.016~\mathrm{mag\,yr}^{-1}$ superimposed on a low-amplitude oscillation, compatible with an orbital period of 5.54\,yr.  
 
Comparison between the whole object brightening and the stellar core indicates that the nebula continued to brighten faster than the stellar core from 1963 to 1999. This means that the circumstellar extinction remained very high for decades after 1940.

\subsection{Magnitudes of the stellar core in the time-frame 1999--2026}

Around 1999, there was a sharp jump of about 3.2 times in the rate of brightness increase for the $V$-magnitude of the whole object (Figure \ref{Fig:Vall}). Unlike the 1940 event, this phenomenon occurred without a change in the nebular brightness \citep{damineli19,mehner19}. Subsequent studies \citep{damineli24}, indicated that this phenomenon was due to the dissipation of a coronagraphic type structure located in our direction at approximately $900-1700$\,AU from the central system, in line with the suggestion of \citet{hillier92}. Because of this asymmetry, the nebula did not reflect the observed increase in brightness, which was restricted exclusively to our direction, continuing to increase slowly in brightness as it did in the 1945--1999 interval. The observed brightness increase rate for the object as a whole,  {$\delta V = -0.051\,\mathrm{mag\,yr}^{-1}$}, is due to the brightness increase of the stellar core, as confirmed by spatially resolved photometry {with} the HST \citep{davidson99}.

The brightening of the stellar core in the time window 1999--2026 jumped by $20 \times$ as compared to that of 1963--1999. After 2010, the stellar core became brighter than the surrounding nebula; however, the contrast remains insufficient to disentangle the core from the nebular emission in ground-based images without space-based imaging. 

\begin{figure}[!ht]
\centering
\includegraphics[width=1\linewidth,angle=0,viewport=0bp 0bp 480bp 460bp, clip]{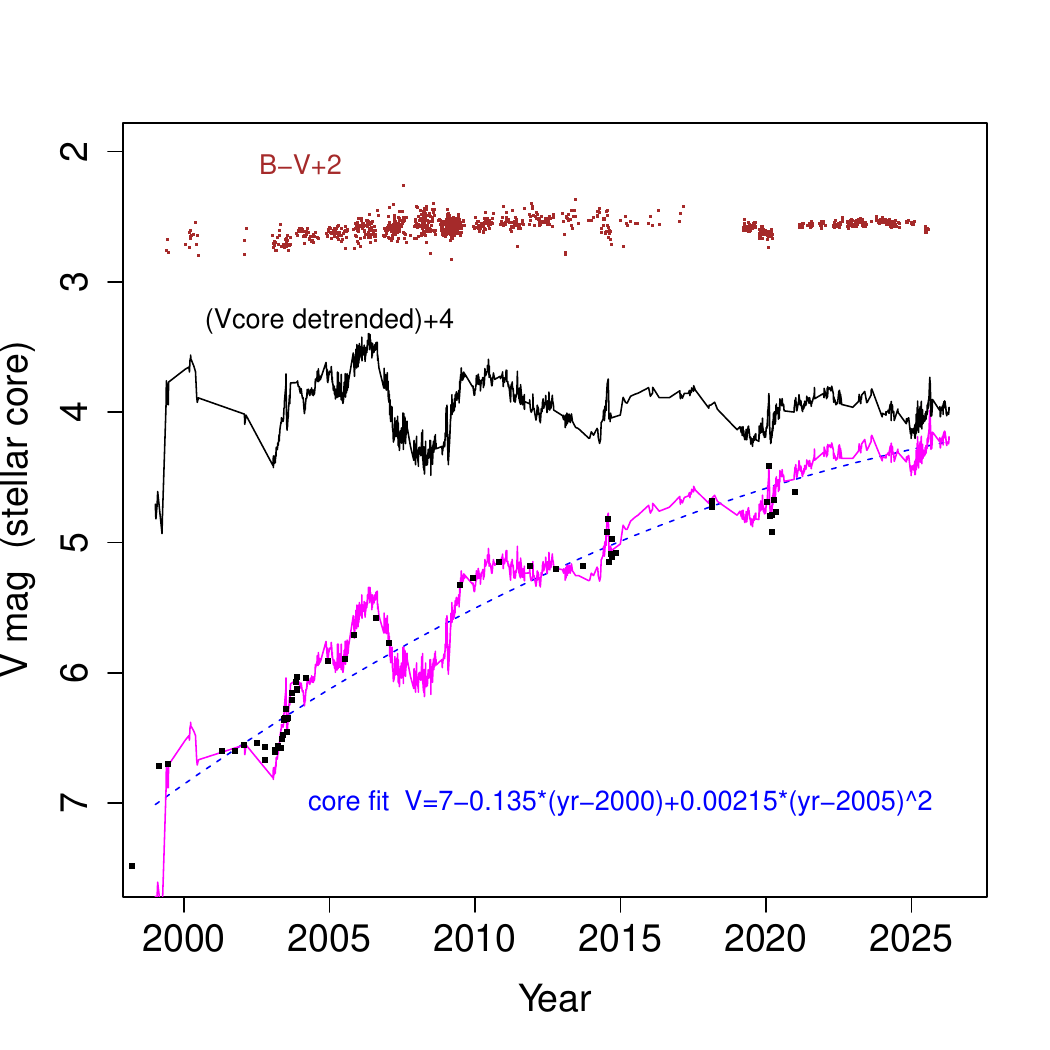}
\caption{Light curve in the $V$-band of the stellar core extracted from ground{-based} measurements (magenta line at bottom) and calibrated using HST/ACS/HRC images and HST/STIS spectra (black points). The dashed line is a 2nd-degree polynomial fit of the brightening. The de-trended light curve (black solid line), after subtracting the long-term fit, is indicated as the intermediate curve. At the top, brown dots represent the {$B-V$} color indices measured for the stellar core in the ground-based light curves; the points were vertically shifted by $+2$\,mag.} 
\label{Fig:fit}
\end{figure}

Unfortunately, visits by the HST/ACS/HRC for {photometric separation of the stellar core from the nebular contribution were sparse} and restricted to the time-frame 1999--2015. \citet{damineli19} developed a simple procedure to use the spatial resolution power of the HST to obtain precise information about the stellar core in the $V$ band from ground-based photometry, which is much more frequent and monitored over a much longer period. In Fig.~\ref{Fig:fit}, we show how this method allowed us to extract the $V$-band magnitude of the stellar core between the years 1999 and 2020. In Fig.~\ref{Fig:Vall}, the filled and empty circles show the values obtained in narrow bands in the STIS/STIS spectra and in the ACS images in the F550M filter. For narrow-band spectrophotometry, we measured the flux in the stellar continuum of HST/STIS spectra at $5495\pm2.5$\,{\AA} for the $V$-band and at $4405\pm2.5$\,{\AA} for the $B$-band. The narrow {spectral range} was adopted to avoid contamination of the spectral features by the continuum.

The sum of the nebular and stellar core flux components coincides with spatially unresolved ground-based photometry. This makes it possible to estimate the magnitude of the nebular component (gray triangles in Fig.~\ref{Fig:Vall}) for all long-term series of the object as a whole (core + nebula) and, therefore, separate the component associated with the stellar core. We projected the same brightening of the nebula from 1999 to the present day. The fit - almost a constant - is represented by a straight, slightly inclined line in Fig.~\ref{Fig:Vall} :
%$\mathrm{V_{nebula}} = 5.5 - 0.017\,\mathrm{(yr-2003.87)}$.    (eq.~4)\\
\begin{equation}
V_{neb} = 5.5 - 0.017\,(\mathrm{yr}-2003.87). 
\end{equation}

The $B$-band magnitude of the nebula ($\approx 6.0$) could not be derived with the same precision as the $V$-band magnitude due to the lack of a corresponding filter on the HST/ACS/HRC camera. The $B$-band magnitude of the stellar core is recovered using just the spectrophotometric flux of the HST/STIS spectra.

\subsection{The orbital light curve}

Applying this method to ground photometry between 1999 and 2026 (Fig.~\ref{Fig:fit}) highlights the behavior of the stellar core (magenta). The black dots on the magenta curve show excellent agreement between the predicted core magnitudes obtained from ground-based observations (magenta) and those spatially resolved, obtained with the HST/STIS (black dots). Note that after subtracting the nebular component ($V \sim 5.5$), the light curve reveals a much sharper orbital oscillation than in the ground light curve, where the two components are added.

\begin{figure}[!ht]
\centering
\includegraphics[width=1\linewidth,angle=0,viewport=0bp 0bp 900bp 830bp, clip]{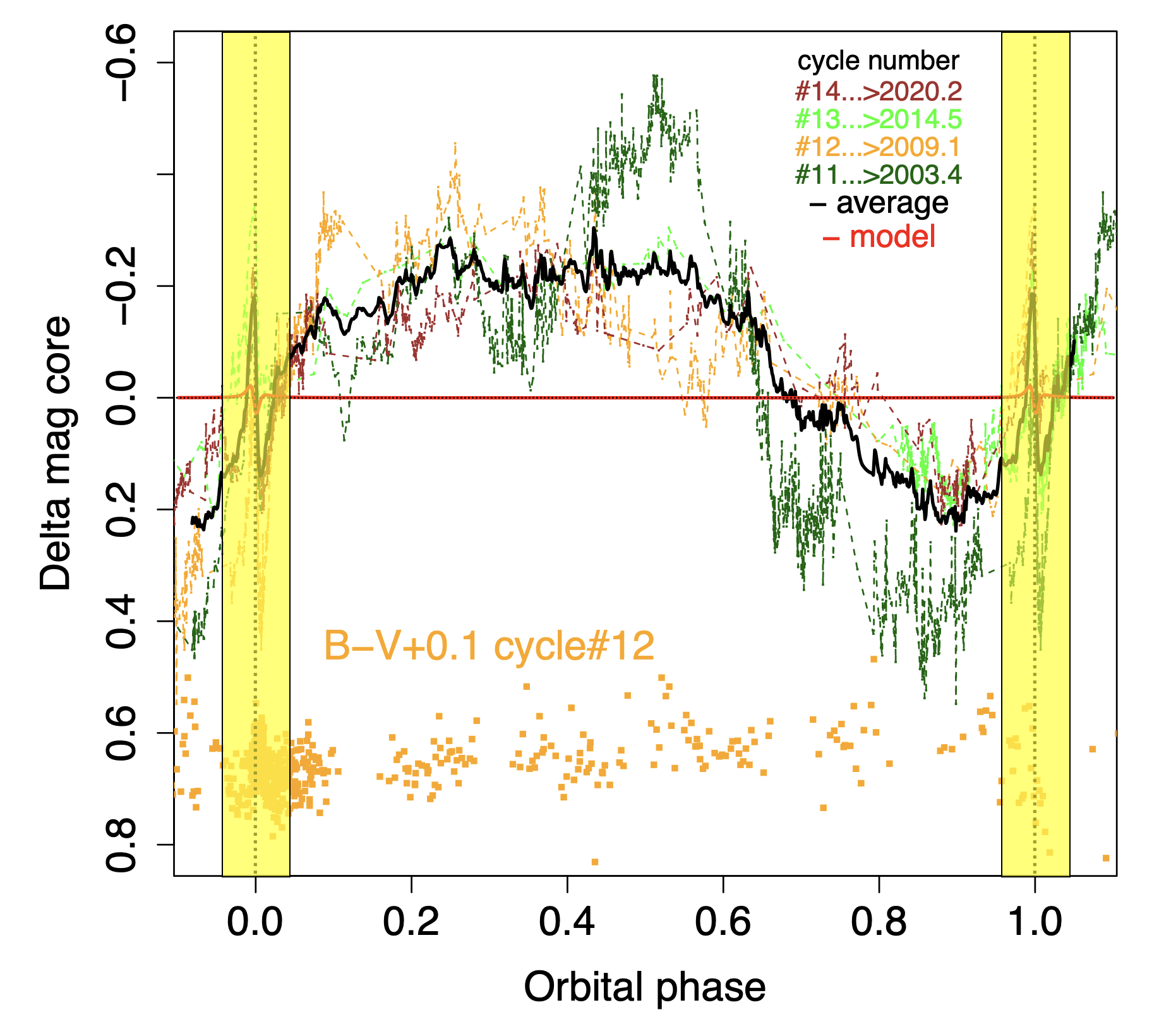}
\caption{Average orbital light curve of the stellar core (black) over the last four orbital cycles (colors).  Model for ellipsoidal variations is shown in red. Yellow boxes show the phases when the Roche lobe radius is smaller than the primary star radius. 
A distinct peak is seen just before periastron (JD\,2460922.03), followed by a local minimum shortly thereafter. The orange points represent the $B-V$ color indices of the stellar core for cycle \#12: almost constant over the orbital cycle (see Sect.~\ref{subsec:model}).}
\label{Fig:orbital}
\end{figure}

We performed a second-order polynomial fit, shown in the lower part of Fig.~\ref{Fig:fit}, so that it could be subtracted from the data to measure the amplitude of the orbital oscillation. 
\begin{equation}
V_{fit} = 7.0 - 0.135\,(\mathrm{yr}-2000) + 0.00215\,(\mathrm{yr}-2005)^2.   
\end{equation}

After that, we removed the secular increase (\emph{trend} - Fig.~\ref{Fig:fit}). Since the increase in brightness of the stellar core is due to the gradual decrease in extinction in the line of sight, the equation shown in that Figure is dominated by extinction in the $V$-band. Note that the extinction decreased at a faster rate in the first three orbital cycles (from 2000) and became milder in the last two. The $V$-magnitude is composed of $V_0=0.91$ \citep{hillier01} plus the foreground extinction, $A_{V_{fg}}=2.45$\,mag \citep{damineli19}, plus the extinction of dissipating circumstellar dust:
%$V=V_0+ AV_{foreg.} +AV_{circumst.}$.         (eq.~6)\\
\begin{equation}
V=V_0 + A_{V_{fg}} + A_{V_{circum}}.
\end{equation}

We averaged the four orbital cycles in Fig.~\ref{Fig:fit} to build the orbital light curve shown in Fig.~\ref{Fig:orbital}, which suggests ellipsoidal deformation of the primary star, since the $B-V$ color index is constant and compatible with the reddened primary star. However, uncorrelated cycle-to-cycle variations indicate a major contribution to the light curve from the matter ejected by the primary.  The large fluctuations of the orbital light curve from one cycle to another also indicate that its origin is not the ellipsoidal deformation of the primary.

\subsection{The stellar core magnitudes in the whole window 1963--2026}

\begin{figure}[!ht]
\centering
{\includegraphics[width=0.9\linewidth,angle=0,viewport=10bp 20bp 835bp 500bp, clip]{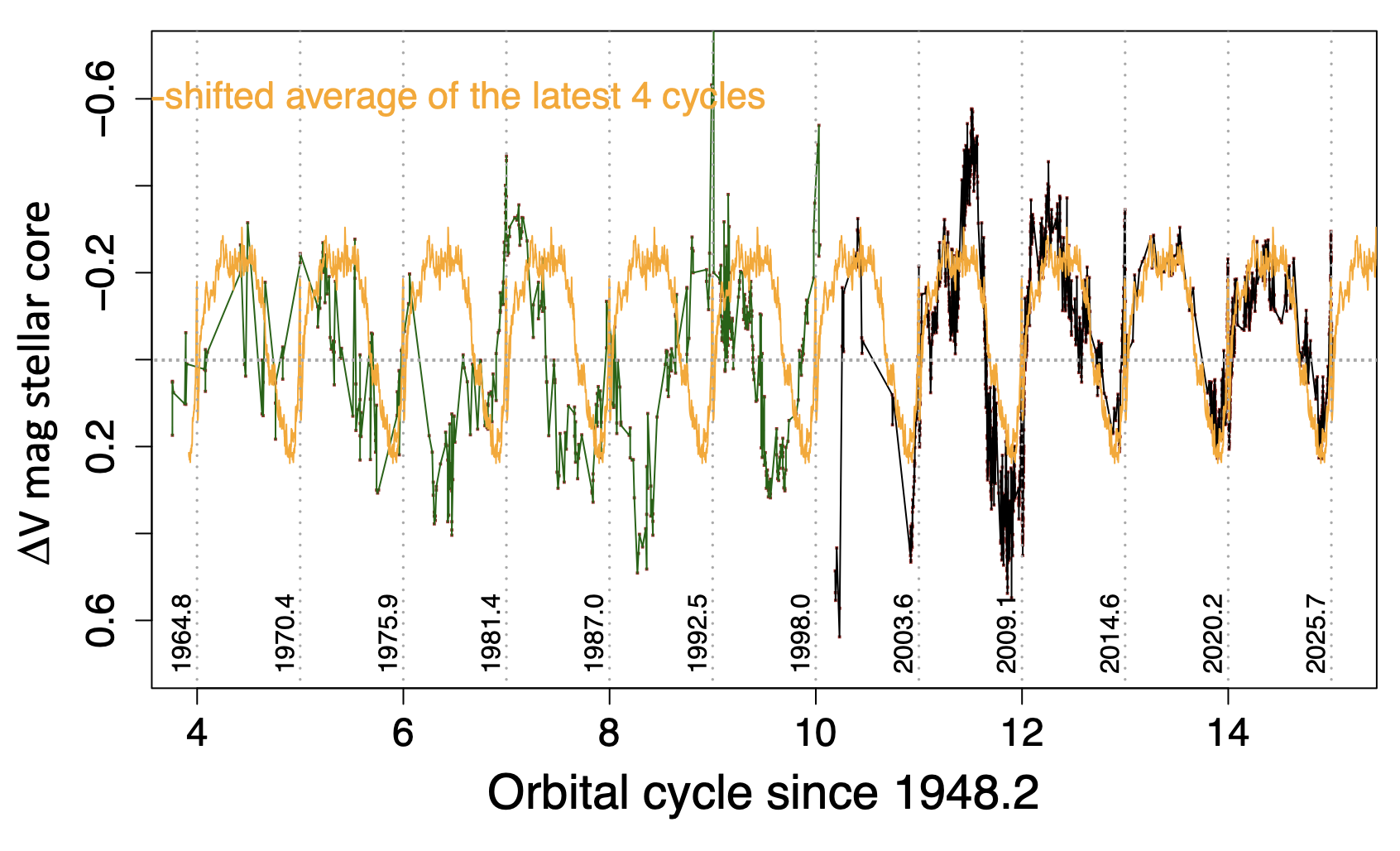}}
\caption{Normalized light curve of the stellar core (extracted from measurements of the whole object) from 1963 on. The average curve of the latest four cycles (orange) is superimposed on the entire normalized light curve (dark green before 1999, plus black afterwards). Cycle numbers are referred to the first low-excitation event reported by \citet{gaviola53}.} 
\label{Fig:Vlong}
\end{figure}

We combined the normalized $V_{core}$ magnitude in the two windows (dark green and black), as shown in Fig.~\ref{Fig:Vlong}. The orbital light curve was shifted in phase all the way up to 1963 and is represented as an orange line in Fig.\ref{Fig:Vlong}, to see the differences between the original light curve and the average. Despite the non-uniform photometric coverage, the presence of significant observational gaps, and potential calibration uncertainties, the photometric oscillations exhibit a nearly constant amplitude over more than six decades of observations.

In several orbital cycles, the photometric peak shows up slightly before the predicted epochs of periastron passage. This photometric behavior suggests that the ground-based light curve exhibits clear signatures of orbital motion since at least 1963.

\subsection{The pre-periastron photometric peak}
\label{peak}

\begin{figure}[!ht]
\centering
{\includegraphics[width=1\linewidth,angle=0,viewport=30bp 20bp 1345bp 750bp, clip]{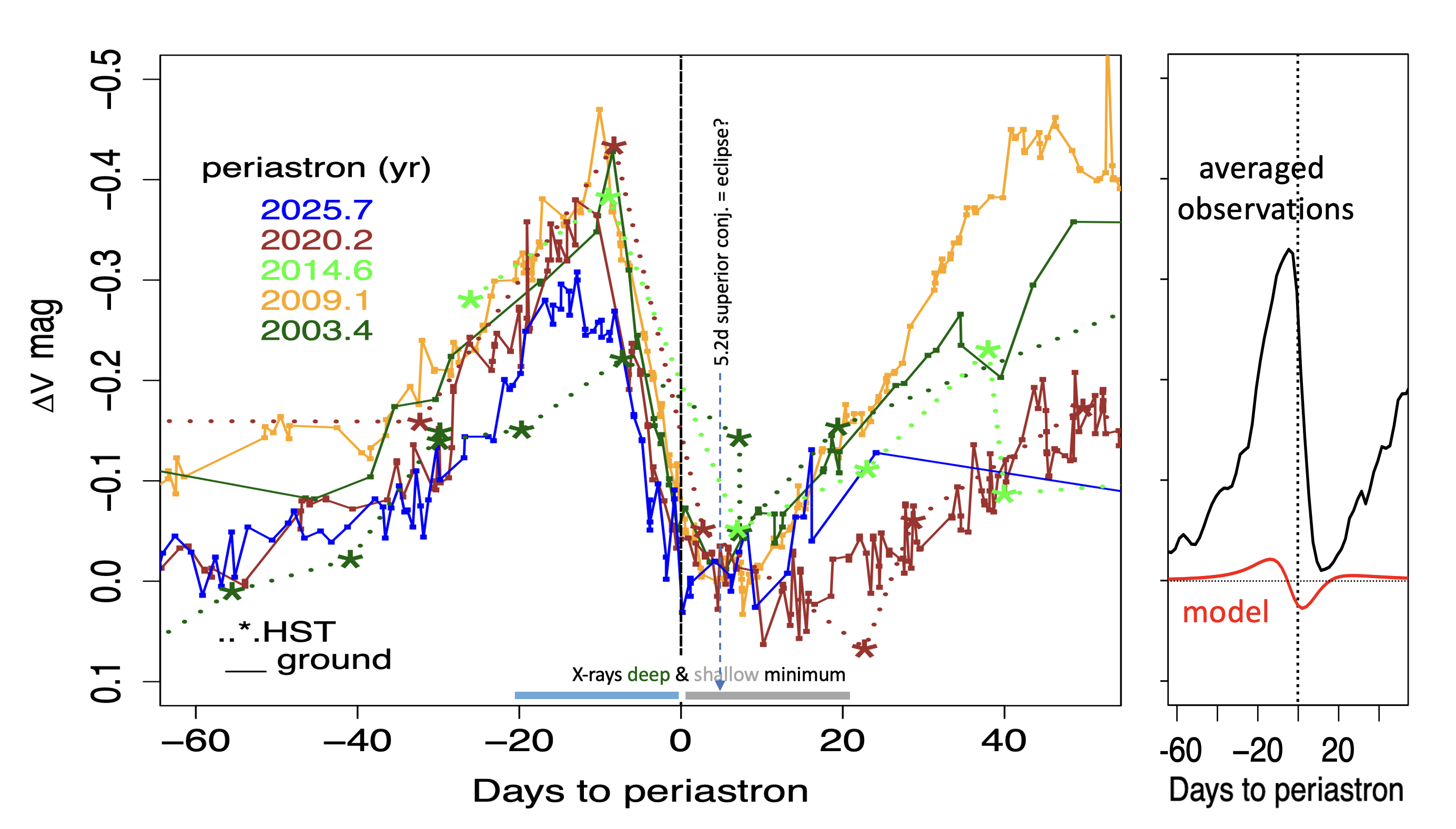}}
\caption{The photometric peak at pre-periastron phases. The light curves are shifted to $\Delta V=0$ at the superior conjunction of the secondary star. Ground-based observations are shown as solid lines,  HST/STIS measurements are shown as dotted lines with polygons ($\ast$). The post-periastron minimum is centered on the date of the superior conjunction of the secondary star ($T_0 + 5.2$ days). The right panel (compressed version of the left panel) shows that the observed variations (black line) are $\sim 10 \times$ larger than the modeled ellipsoidal amplitude (red).}
\label{Fig:peak}
\end{figure}

The photometric peak is a distinctive feature in the pre-periastron period. Fig.~\ref{Fig:peak} shows the light curves shifted such that $\Delta V=0$ at the epoch of superior conjunction of the secondary star. It is positioned in correspondence with the \emph{shallow minimum} observed in X-rays, which is interpreted as an eclipse. The \emph{deep minimum} occurs just before the periastron (-20 to 0 days) and is interpreted to be due to a wind-wind collision occurring at non-terminal speed  \citep{espinoza22}. The good correspondence of the HST curves with those from ground-based observations reinforces that the calibration to recover the nebular brightness is reliable. The maximum intensity of the pre-periastron peak decreases with time: $\Delta V \approx 0.45$ in 2003.4 and 2009.1 to $\Delta V \approx 0.3$ in 2020.2 and 2025.7. The nature of this peak is uncertain. It may be when the primary star has reached its largest Roche lobe overfilling near periastron. Note that it always occurs at the same orbital phase, and the {$B-V$} color index of the stellar core remained roughly constant. The flow of gasses towards the secondary was also suggested by \citet{soker01} and \citet{kashi19}.

The presence of a photometric minimum shortly after the periastron suggests an eclipse-like event. The minimum is centered near the epoch of superior conjunction of the secondary star, which occurs $\approx 5.2$\,d after the periastron. However, the observed depth of the minimum ($\Delta V \sim 0.1$\,mag) is significantly larger than expected from the eclipse of the secondary star alone. Given the luminosity ratio of the two components, $L_1/L_2 > 33$ (Weigelt 2026, in preparation), complete occultation of the secondary would produce a brightness decrease of only $\Delta V \sim 0.03$\,mag. This discrepancy suggests that the eclipsed source is not the secondary star itself but rather the additional luminosity associated with the material accreting onto it.

\subsection{Is the orbital brightness modulation due to ellipsoidal deformation of the primary?}
\label{subsec:model}

To investigate whether the observed orbital modulation in the $V$-band light curve can be attributed to the geometric distortion of the primary star, we computed a synthetic light curve using the Wilson-Devinney (WD) code \citep{wilson1971}. The model was constrained using a priori orbital parameters derived from the radial velocity modeling by Jablonski et al. (in prep.) and Navarete et al. (in prep.)The adopted Keplerian orbital elements are the orbital period $P = 2022.35$\,days, periastron passage time $T_0 = \rm JD~ 2,460,922.03$, periastron argument $\omega = 246.1^\circ$, eccentricity $e = 0.862$, orbital inclination $i = 134^\circ$ and inverse mass ratio $ M_1/M_2 = 1.1$.

In our WD modeling, we assumed that the primary star completely fills its Roche lobe exactly at the periastron, representing the phase of maximum tidal deformation. The absolute dimensions and effective temperatures adopted for the components were $R_1 = 188.5 R_\odot$ and $T_1 = 18000$\,K for the primary and $R_2 = 30 R_\odot$ and $T_2 = 39000$\,K for the secondary. The resulting synthetic $V$-band light curve is virtually flat throughout most of the orbital cycle. The only notable photometric variations occur near the periastron passage (between phases $\phi = 0.96$ and $\phi = 1.04$). During this brief interval, the model predicts a slight brightness increase of $\approx0.02$\,mag, immediately followed by a decrease of $\approx 0.05$\,mag, strictly driven by the extreme ellipsoidal deformation of the primary star in the strong gravitational gradient. However, interferometric measurements from \citet{weigelt21} indicate a much larger effective radius for the primary $R_1 \approx 700\, R_\odot$, which vastly exceeds the Roche lobe of the actual orbit and likely represents the optically thick wind pseudo-photosphere. To test the photometric impact of such a large radius within the WD hydrostatic framework, we artificially scaled the orbital separation until the primary's Roche lobe could accommodate this $700\, R_\odot$ radius, maintaining the Roche lobe filling condition at the periastron. As expected from the scale-free nature of Roche geometry for a fixed mass ratio, the resulting synthetic light curve remained virtually unchanged. This test demonstrates that geometric ellipsoidal variations, regardless of the assumed stellar scale, are fundamentally insufficient to reproduce the large photometric amplitudes observed during the periastron passage, as seen in Fig.~\ref{Fig:orbital} and the right panel of Fig.~\ref{Fig:peak} with the red curve.

\subsection{Roche Lobe overflow of the primary star?}
\label{rolof}

In order to obtain further information on the role of the ellipsoidal deformation of the primary star, we used the \citet{eggleton83} approximation:

\begin{equation}
R_{RL_1} = \frac{0.49 Q^{2/3}}{0.6 Q^{2/3} + \ln(1 + Q^{1/3})}\, a\\
\label{eq7}
\end{equation}

where $Q = M_1/M_2$ and  a is the instantaneous distance between the stars.

Due to the high eccentricity, the Roche lobe radius of the primary star shrinks by a factor of $\approx 20$ from the apastron to the periastron. This suggests that at some point the stellar radius overflows the Roche lobe. If we adopt the radius measured by \citet{weigelt21} $R_1 \approx 700\, R_\odot$, this occurs 75 days before the periastron. During the time-frame of 150 days there is matter thrown outside the primary star Roche lobe, which is represented by a yellow vertical strip in Fig.~\ref{Fig:orbital}. 
Not by coincidence, at phase 0.962 (or -0.038), the orbital light curve starts to increase. This is because the expanding ejected matter reflects the light of the primary star, which has the same $B-V$ color index as that of the primary. The continuously expanding cloud of ejected material explains why the orbital light curve continues to increase for the phases subsequent to periastron.

The Keplerian solution of the radial velocity curve indicates that the secondary star reaches the superior conjunction at $T_0 + 5.2$ days (see Fig.~\ref{Fig:peak}). At this phase, the material ejected towards the secondary star is eclipsed, producing a local minimum. In this way, the photometric peak may not be a sudden ejection but just the left edge of the post-periastron eclipse or a combination of the two.

\subsection{The size of the evaporating dust grains to our line of sight}

\begin{figure}[!ht]
\centering
{\includegraphics[width=1\linewidth,angle=0,viewport=15bp 0bp 1050bp 910bp, clip]{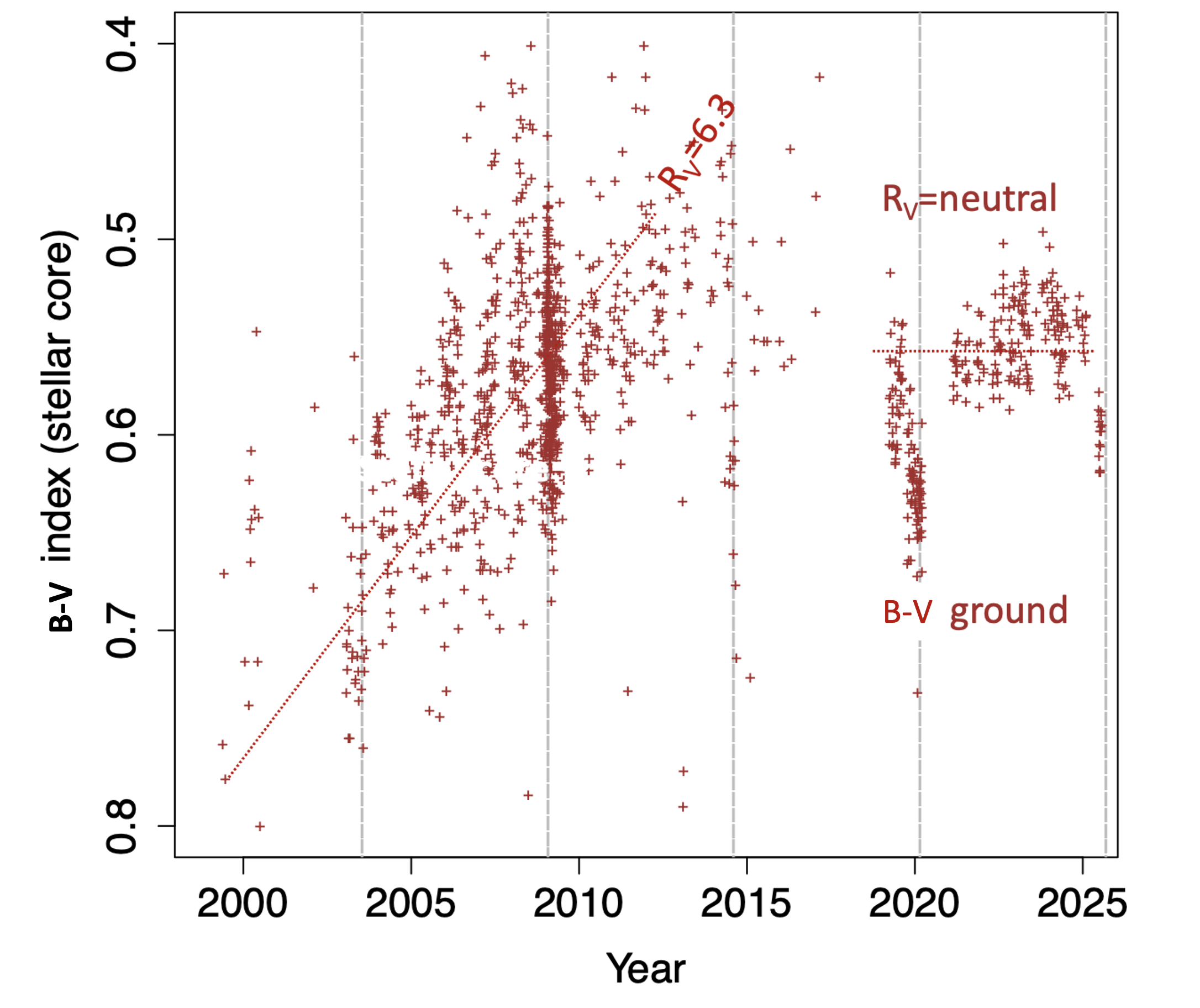}}
\caption{Evolution of stellar core color index $B-V$ measured from the ground,  calibrated with HST/STIS. In the interval 1999--2012, while the star increased $\Delta V = -1.55$\,mag in brightness, the $B-V$ color index decreased $\Delta V = -0.25$\,mag. The color indices remained constant in the last two cycles (from 2015 onwards), indicating that the slow increase in $V$-band brightness presently is due to a slowly decreasing neutral extinction. The $U-B$ color index remained roughly constant since 1999 (not shown here), indicating that the average grain size has been larger than the corresponding $B$-band wavelength.} 
\label{Fig:RV}
\end{figure}

The observed bluening – Fig.~\ref{Fig:RV} – in the {$B-V$} color index of the stellar core (defined as $B - V = B_{core} - V_{core}$  of $\eta$ Car was $\Delta(B-V) = -0.25$ while the star increased in brightness by $\Delta V \sim -1.55$\,mag) implies a reddening law $R_V=A_V/E(B-V)=6.2$. The grain models for that period \citep{draine03,rumpl77} indicate that the relationship between the effective diameter of the grain $a_{\rm eff}$ and the reddening law ($R_V$) is:
\begin{equation}    
a_{\mathrm{eff}} = 0.1 (R_V/3.1) \micron \approx 0.2 \micron
\label{eq8}
\end{equation}

The maximum size of grains is:
\begin{equation}    
a_{\mathrm{max}} = 0.25 (R_V/3.1)^{1/\alpha} \micron \approx 1 \micron    \label{eq9}
\end{equation}

Using $\alpha = 0.7$, the minimum grain size is:
\begin{equation}    
a_{\mathrm{min}} = a_{\mathrm{eff}}^2 / a_{\mathrm{max}} \approx 0.04 \micron
\label{eq10}
\end{equation}

\citet{hillier01} gives a value of $R_V = 5.0$ with two additional magnitudes of neutral extinction. The bluening of the $B-V$ color index in this phase (1999--2012) indicates that the size of the typical dust grains was smaller than the wavelength of visible light. This is corroborated by the constancy of the $U-B$ color index (not shown), suggesting a neutral extinction for those wavelengths. During the last two orbital cycles, the star continued to increase in brightness (at a slower pace), and now all color indices remain constant, indicating that the dust grains producing the neutral extinction are large.  The removal of the smaller grains left only the large grains ($\geq$1{\micron} that are still in the process of destruction. \\
As for comparison, \citet{morris17} modeled the infrared SED of the entire Homunculus and reported that the main components of the dust grains contributing to the total dust mass and extinction are olivine and metal-rich silicates of diameter $a_{min} \geq 3.2 \mu$m).

\section{Discussion and Conclusions }
\label{discussions}

The $V$-band light curve of \ec after 1940 cannot be modeled when using the ground-based photometry alone because the stellar core is mixed with the nebular emission. Although the nebula evolves slowly, the stellar core varies in the short- and long-term in magnitude and color index.

Coeval observations from the ground and HST in the time frame (1999--2020) enabled the modeling of the ground-based nebular flux, which is almost constant ($ V \approx 5.5$). The 1999 stellar core magnitudes were bridged with those of 1940-46, when the core was much brighter than the nebula and could be easily measured by differential photometry. The brightness of the nebula in 1999 could be in error because the stellar core was too faint.The most accurate values are those after that epoch, since they were calibrated using HST/ACS/HRC and HST/STIS.
 
The normalized light curve presented in Fig.~\ref{Fig:Vlong} shows a sinusoidal light curve, especially after 1999. It suggests an orbital deformation of the primary star, constrained within its Roche lobe, with variable size in the very elliptical orbit ($e = 0.86$). However, this is not the case, since our synthetic $V$-band light curve (Fig.~\ref{Fig:orbital}) exhibits negligible photometric oscillation in the intermediate orbital phases. Near the periastron passage, the modeled light curve reveals only a slight peak and a subsequent minor dip, with amplitudes far smaller than those actually observed. This predicted dip (shown in the red curve in the right panel of Fig.~\ref{Fig:peak}) arises strictly from the extreme ellipsoidal deformation of the primary star and coincides temporally with the superior conjunction of the secondary. However, the observed depth of this minimum vastly exceeds the predicted geometric effect, providing strong evidence for the occultation of an extended luminous source. We propose that this extra luminosity associated with the secondary arises from an active accretion flow. Indeed, interferometric observations reveal a significant amount of mass distributed around the primary at these orbital phases \citep{weigelt21}. The large amplitude of the observed orbital light curve suggests periodic Roche lobe overflow, leading to episodic mass transfer towards the companion. In this scenario, mass ejection from the primary initiates before the periastron ($\phi \approx 0.96$) and contributes to the excessive brightness of the broad-band until $\phi \approx 0.6$. As this material dissipates, the light curve reaches a minimum near $\phi = 0.85$, setting the stage for a new mass ejection cycle, as can be seen in Fig.\ref{Fig:orbital}. Roughly two months before periastron, a photometric peak starts to appear, reaching its maximum at  $T_0-20$\,days. At this point, the primary star eclipses part of the spilled material towards the companion, producing a nearly two-month-long eclipse. The deepest part of the minimum is centered on  $T_0+5.2$\,days, in exact coincidence with the superior conjunction of the secondary and derived from the radial velocity curve (Jablonski et al., in prep.). This minimum is correlated with the \emph{shallow minimum} observed in X-rays, which is also interpreted as an eclipse \citep{espinoza22,hamaguchi14}. The \emph{shallow minimum} in X-rays coincides with the epoch of periastron and is interpreted as due to wind-wind collision at a speed lower than terminal, consistent with a hardness-ratio smaller than in other orbital phases.

The evolution of the $B-V$ color index shows that in the time frame \textcolor{blue}{1999--2012, relatively small grains in addition to large ones} were in the process of destruction in the \emph{occulter} material in our line-of-sight. The reddening law $R_V = A_V/E(B-V)=6.2$ may be overestimated, since part of the $V$-band magnitude change could be due to gray extinction. In that case, the grain size in equations~\eqref{eq8}--\eqref{eq10} would be larger. In the last two orbital periods (after 2015), only very large grains survived, producing neutral reddening, and continued to be destroyed. Except for the long-term evolution of the $B-V$ color index, it is amazingly constant over an orbital cycle, indicating that all orbital variations are produced by material reflecting the light of the primary star.

An event of particular importance occurred around 1999 that led to the destruction of a large amount of dust in our \textcolor{blue}{ direction and towards the Weigelt clumps \citep{damineli24} } world and has continued until the present time.  The initial extinction was $\approx 6-7$\,mag, and there is still $\approx 1$\,mag to be cleaned in the \emph{coronagraphic} occulter. The stellar core will end up at $V \sim 4.05$\,mag when the circumstellar material to our line-of-sight will be cleaned up, which should occur around $\approx$2031.  After allowing for $V \sim 0.5$ for the intervening extinction of Homunculus, the final brightness would be $V \sim 3.5$. If we add the extinction as composed also by ISM $A_V \sim 1.12$ and intra-cluster $A_V \sim 0.84$ \citep{damineli19}, the derived unobscured magnitude would be $V_0 \sim 1$\,mag in excellent agreement with that reported by \citep{hillier01} obtained by a different approach. In a few years, the nebular brightness will become much smaller than that of the stellar core, substantially decreasing the impact of the nebular contamination to the stellar core in ground-based measurements.
 
     The final apparent magnitude  $V \sim 3.5$ will be very close to the pre-eruption brightness around 1600-1700 \citep{frew04}. This close coincidence is a little uncomfortable if we adopt the scenario of a binary merger in the Great Eruption \citep{smith18c, hirai21}  since the internal mixing of the resultant object would increase its luminosity by $\approx 1$\,mag \citep{schneider19}. Another problem with the binary merger scenario is that it would result in a fast-rotating remnant. High rotation was not shown directly yet, as it would cause S~Doradus instabilities, as is the case for other LBVs \citep{Groh09}.

 If our scenario of periodic mass ejections at periastron is right, the impact on preriapsis precession would be important. In addition to the large mass loss rate of the primary, the impact on the orbit of the system could be important.

\section{Data Availability}
The Tables are available in the online journal and in CDS/Vizier.

\section{Acknowledgments}
AD thanks to FAPESP (grants 2011/51680-6 and 2024/16086-6) and CNPq grant 304346/2023-3.
R. Capobiango thanks FAPESP for the fellowship 2025/08626-3. 
FN thanks the support of CNPq (grant 303093/2025-0). LAA thanks the support of CNPq (grant 313032/2025-4). FJ acknowledges the Brazilian Ministry of Science, Technology and Innovation (MCTI) and the Brazilian Space Agency (AEB), for support under PO 20VB.0009.
EM acknowledges funding from FAPEMIG under project number APQ-02493-22 and the CNPq grant 309829/2022-4.
We thank the observatories: VSN/OALP, HSH/Casleo, OPD/LNA, and OC61/Mount John operated by AAVSO, for enabling the observaations.

\bibliographystyle{aasjournal}
\bibliography{refs} 

\appendix

\section{Full photometric tables = to appear in CDS as machine readable tables.}
\label{appendix_tables}

\begin{table}[!ht]
\centering
\caption{$V$-band photometry taken at Pico dos Dias Observatory in Aug/2025.}
\label{table_S50}
\begin{tabular}{ccc}
\hline
\hline
JD & $V$ & $\sigma_V$ \\
\hline
2460890.251 & 3.945 & 0.07 \\
2460896.301 & 3.895 & 0.06 \\
2460897.280 & 3.853 & 0.05 \\
2460898.270 & 3.876 & 0.05 \\
2460899.252 & 3.876 & 0.05 \\
2460900.255 & 3.832 & 0.05 \\
2460902.288 & 3.839 & 0.04 \\
2460903.275 & 3.810 & 0.04 \\
2460905.310 & 3.810 & 0.04 \\
2460906.272 & 3.741 & 0.07 \\
2460907.271 & 3.733 & 0.07 \\
2460909.270 & 3.770 & 0.09 \\
2460910.265 & 3.784 & 0.06 \\
\hline
\end{tabular}

\vspace{0.5em}
\textbf{Notes:} TELESCOPE: 0.05-m (S50 ZWO)\\
OBSERVING and PROCESSING: Rodrigo Capobiango\\
Magnitudes are in the La Plata system
\end{table}

\begin{table}[!ht]
\centering
\caption{$BVRI$-band photometry taken at La Plata Jan-Jun 2017.}
\label{table_VSN}
\begin{tabular}{ccccccccc}
\hline
\hline
JD & $B$ & $\sigma_B$ & $V$ & $\sigma_V$ & $R$ & $\sigma_R$ & $I$ & $\sigma_I$ \\
\hline
2457768.6038 & 4.962 & 0.036 & 4.234 & 0.009 & 3.427 & 0.043 & 2.968 & 0.014 \\
2457775.5750 & 4.955 & 0.036 & 4.274 & 0.009 & 3.41  & 0.018 & 2.99  & 0.018 \\
2457775.5879 & 4.942 & 0.028 & 4.269 & 0.013 & --    & --    & --    & --    \\
2457800.5529 & --    & --    & 4.268 & 0.012 & 3.405 & 0.020 & 2.973 & 0.018 \\
2457825.4849 & --    & --    & 4.239 & 0.019 & 3.426 & 0.015 & 2.951 & 0.020 \\
2457875.4857 & 4.946 & 0.036 & 4.231 & 0.008 & 3.383 & 0.010 & 2.924 & 0.009 \\
2457889.5065 & --    & --    & 4.213 & 0.004 & 3.396 & 0.026 & --    & --    \\
2457895.5114 & 4.957 & 0.028 & 4.240 & 0.006 & 3.381 & 0.022 & 2.928 & 0.027 \\
2457903.4870 & --    & --    & 4.213 & 0.021 & 3.376 & 0.012 & --    & --    \\
2457914.5652 & 4.962 & 0.013 & 4.221 & 0.008 & 3.370 & 0.008 & 2.918 & 0.009 \\
2457916.4312 & 4.898 & 0.027 & 4.196 & 0.009 & 3.368 & 0.013 & 2.909 & 0.015 \\
\hline
\end{tabular}

\vspace{0.5em}
\textbf{Notes:} TELESCOPE: 0.80-m Virpi S. Niemela (VSN)\\
CAMERA: CCD, Photometrics STAR I\\
OBSERVATORY: Observatorio Astronomico de La Plata, La Plata, Argentina\\
OBSERVING and PROCESSING: E. Fernandez-Lajus, et al.\\
The magnitudes are in the La Plata system.
\end{table}

\begin{center}
\begin{longtable}{rcccccccccc}
\caption{$UBVRI$-band photometry taken at Casleo 2019-2025.}
\label{table_HSH}\\
\hline\hline
JD & U & $\sigma$U & B & $\sigma$B & V & $\sigma$V & R & $\sigma$R & I & $\sigma$I \\
\hline
\endfirsthead

\multicolumn{11}{c}%
{{\tablename\ \thetable{} -- continued from previous page}} \\
\hline 
\hline
JD & $U$ & $\sigma_U$ & $B$ & $\sigma_B$ & $V$ & $\sigma_V$ & $R$ & $\sigma_R$ & $I$ & $\sigma_I$ \\
\hline 
\endhead

\hline \multicolumn{11}{r}{{Continued on next page}} \\
\endfoot

\hline \hline
\endlastfoot

2458572.5273	&	--	&	--	&	5.002	&	0.015	&	4.332	&	0.033	&	3.437	&	0.032	&	3.007	&	0.023	\\
2458574.4904	&	--	&	--	&	4.994	&	0.032	&	4.338	&	0.028	&	3.438	&	0.057	&	3.077	&	0.030	\\
2458575.4879	&	--	&	--	&	5.006	&	0.025	&	4.342	&	0.060	&	3.488	&	0.039	&	3.055	&	0.023	\\
2458576.4895	&	--	&	--	&	4.994	&	0.019	&	4.349	&	0.058	&	3.439	&	0.032	&	3.015	&	0.041	\\
2458582.4811	&	--	&	--	&	5.014	&	0.029	&	4.357	&	0.075	&	3.487	&	0.013	&	3.055	&	0.016	\\
2458583.4881	&	--	&	--	&	--	&	--	&	4.354	&	0.056	&	3.480	&	0.055	&	3.061	&	0.037	\\
2458584.4748	&	--	&	--	&	4.985	&	0.016	&	4.365	&	0.017	&	3.478	&	0.030	&	3.042	&	0.024	\\
2458585.4691	&	--	&	--	&	5.004	&	0.020	&	4.357	&	0.033	&	3.496	&	0.038	&	3.064	&	0.031	\\
2458586.4617	&	--	&	--	&	5.006	&	0.033	&	4.367	&	0.057	&	3.471	&	0.047	&	3.047	&	0.033	\\
2458604.4523	&	--	&	--	&	5.011	&	0.016	&	4.339	&	0.040	&	3.481	&	0.056	&	3.062	&	0.019	\\
2458605.4864	&	--	&	--	&	5.015	&	0.008	&	4.347	&	0.023	&	3.490	&	0.018	&	3.041	&	0.008	\\
2458606.4554	&	--	&	--	&	4.996	&	0.044	&	4.342	&	0.083	&	3.431	&	0.054	&	3.046	&	0.043	\\
2458607.5676	&	--	&	--	&	5.015	&	0.010	&	4.354	&	0.033	&	3.475	&	0.041	&	3.018	&	0.016	\\
2458608.4633	&	--	&	--	&	5.018	&	0.028	&	4.355	&	0.025	&	3.464	&	0.028	&	3.019	&	0.010	\\
2458617.4465	&	--	&	--	&	4.989	&	0.022	&	4.319	&	0.060	&	3.468	&	0.052	&	3.047	&	0.024	\\
2458628.4561	&	--	&	--	&	4.956	&	0.033	&	4.324	&	0.061	&	3.458	&	0.041	&	3.000	&	0.040	\\
2458636.4804	&	--	&	--	&	4.982	&	0.030	&	4.306	&	0.048	&	3.435	&	0.067	&	3.008	&	0.024	\\
2458638.4366	&	--	&	--	&	4.979	&	0.021	&	4.304	&	0.075	&	3.461	&	0.061	&	3.014	&	0.028	\\
2458639.4375	&	--	&	--	&	4.971	&	0.027	&	4.326	&	0.055	&	3.450	&	0.036	&	2.986	&	0.030	\\
2458640.4411	&	--	&	--	&	4.996	&	0.012	&	4.324	&	0.033	&	3.460	&	0.033	&	3.004	&	0.014	\\
2458641.4459	&	--	&	--	&	4.995	&	0.020	&	4.340	&	0.024	&	3.461	&	0.023	&	3.011	&	0.012	\\
2458642.4408	&	--	&	--	&	4.998	&	0.020	&	4.323	&	0.060	&	3.459	&	0.043	&	2.988	&	0.029	\\
2458645.4387	&	--	&	--	&	4.996	&	0.011	&	4.330	&	0.042	&	3.477	&	0.021	&	3.029	&	0.011	\\
2458654.4799	&	--	&	--	&	5.015	&	0.016	&	4.353	&	0.066	&	3.493	&	0.025	&	3.060	&	0.025	\\
2458656.4405	&	--	&	--	&	5.019	&	0.021	&	4.346	&	0.056	&	3.493	&	0.030	&	3.036	&	0.016	\\
2458661.4593	&	--	&	--	&	5.020	&	0.015	&	4.346	&	0.034	&	3.481	&	0.017	&	3.048	&	0.014	\\
2458670.4620	&	--	&	--	&	5.015	&	0.018	&	4.363	&	0.046	&	3.475	&	0.039	&	3.029	&	0.013	\\
2458676.4799	&	--	&	--	&	5.026	&	0.011	&	4.361	&	0.084	&	3.499	&	0.058	&	3.021	&	0.026	\\
2458677.4444	&	--	&	--	&	5.024	&	0.036	&	4.371	&	0.034	&	3.497	&	0.034	&	3.035	&	0.022	\\
2458680.4469	&	--	&	--	&	5.031	&	0.017	&	4.379	&	0.051	&	3.481	&	0.029	&	3.037	&	0.023	\\
2458682.4568	&	--	&	--	&	5.040	&	0.035	&	4.385	&	0.062	&	3.489	&	0.071	&	3.035	&	0.044	\\
2458684.4558	&	--	&	--	&	5.030	&	0.009	&	4.364	&	0.047	&	3.492	&	0.046	&	3.051	&	0.026	\\
2458689.4513	&	--	&	--	&	4.998	&	0.021	&	4.352	&	0.044	&	3.489	&	0.035	&	3.051	&	0.033	\\
2458693.4581	&	--	&	--	&	5.006	&	0.023	&	4.346	&	0.081	&	3.483	&	0.056	&	3.084	&	0.071	\\
2458694.4701	&	--	&	--	&	4.998	&	0.032	&	4.354	&	0.057	&	3.483	&	0.060	&	3.047	&	0.027	\\
2458697.4662	&	--	&	--	&	5.005	&	0.017	&	4.346	&	0.042	&	3.475	&	0.029	&	3.044	&	0.023	\\
2458699.4594	&	--	&	--	&	4.997	&	0.011	&	4.344	&	--	&	3.478	&	0.039	&	--	&	--	\\
2458700.4617	&	--	&	--	&	4.988	&	0.015	&	4.335	&	0.034	&	3.486	&	0.072	&	3.009	&	0.044	\\
2458702.4616	&	--	&	--	&	4.985	&	0.028	&	4.332	&	0.046	&	3.481	&	0.044	&	3.006	&	0.040	\\
2458704.4624	&	--	&	--	&	4.998	&	0.013	&	4.356	&	0.054	&	3.476	&	0.033	&	3.040	&	0.015	\\
2458709.4695	&	--	&	--	&	4.989	&	0.011	&	4.351	&	0.107	&	3.481	&	0.074	&	3.124	&	0.021	\\
2458715.4659	&	--	&	--	&	4.979	&	0.034	&	4.327	&	0.062	&	3.488	&	0.049	&	3.056	&	0.031	\\
2458718.4630	&	--	&	--	&	4.982	&	0.016	&	4.345	&	0.080	&	3.499	&	0.055	&	3.129	&	0.035	\\
2458722.4562	&	--	&	--	&	4.998	&	0.032	&	4.337	&	0.053	&	3.495	&	0.047	&	3.036	&	0.027	\\
2458723.4625	&	--	&	--	&	--	&	--	&	4.350	&	0.025	&	3.444	&	0.011	&	3.060	&	0.017	\\
2458764.8707	&	--	&	--	&	4.982	&	0.017	&	4.274	&	0.056	&	3.423	&	0.034	&	2.990	&	0.019	\\
2458766.8723	&	--	&	--	&	4.996	&	0.013	&	4.326	&	0.057	&	3.434	&	0.044	&	2.962	&	0.023	\\
2458768.8710	&	--	&	--	&	4.983	&	0.012	&	4.296	&	0.033	&	3.432	&	0.031	&	2.975	&	0.017	\\
2458773.8649	&	--	&	--	&	4.980	&	0.016	&	4.286	&	0.056	&	3.450	&	0.043	&	2.966	&	0.024	\\
2458779.8451	&	--	&	--	&	4.972	&	0.014	&	4.304	&	0.038	&	3.423	&	0.034	&	2.951	&	0.019	\\
2458781.8613	&	--	&	--	&	4.971	&	0.008	&	4.302	&	0.036	&	3.447	&	0.018	&	2.971	&	0.015	\\
2458782.8396	&	--	&	--	&	4.977	&	0.019	&	4.270	&	0.064	&	3.426	&	0.053	&	2.970	&	0.035	\\
2458785.8502	&	--	&	--	&	4.960	&	0.013	&	4.260	&	0.038	&	3.433	&	0.038	&	2.921	&	0.017	\\
2458787.8487	&	--	&	--	&	4.973	&	0.020	&	4.298	&	0.080	&	3.410	&	0.038	&	2.987	&	0.028	\\
2458789.8592	&	--	&	--	&	4.975	&	0.014	&	4.295	&	0.083	&	3.414	&	0.064	&	2.976	&	0.030	\\
2458792.8389	&	--	&	--	&	4.962	&	0.024	&	4.299	&	0.076	&	3.437	&	0.040	&	2.927	&	0.034	\\
2458794.8500	&	--	&	--	&	4.959	&	0.012	&	4.303	&	0.081	&	3.375	&	0.043	&	2.931	&	0.033	\\
2458799.8432	&	--	&	--	&	4.952	&	0.016	&	4.271	&	0.038	&	3.398	&	0.048	&	2.946	&	0.018	\\
2458801.8430	&	--	&	--	&	4.937	&	0.019	&	4.283	&	0.051	&	3.368	&	0.035	&	2.902	&	0.019	\\
2458803.8491	&	--	&	--	&	4.946	&	0.009	&	4.257	&	0.031	&	3.390	&	0.047	&	2.900	&	0.015	\\
2458806.8462	&	--	&	--	&	4.934	&	0.014	&	4.228	&	0.049	&	3.373	&	0.044	&	2.878	&	0.022	\\
2458808.8417	&	--	&	--	&	4.941	&	0.011	&	4.261	&	0.053	&	3.380	&	0.041	&	2.913	&	0.014	\\
2458810.8397	&	--	&	--	&	4.951	&	0.011	&	4.279	&	0.045	&	3.367	&	0.035	&	2.905	&	0.017	\\
2458813.8544	&	--	&	--	&	4.970	&	0.007	&	4.281	&	0.036	&	3.391	&	0.049	&	2.915	&	0.027	\\
2458815.8399	&	--	&	--	&	4.971	&	0.015	&	4.281	&	0.054	&	3.395	&	0.025	&	2.928	&	0.017	\\
2458818.8655	&	--	&	--	&	4.976	&	0.020	&	4.309	&	0.036	&	--	&	--	&	2.885	&	0.021	\\
2458820.7971	&	--	&	--	&	4.983	&	0.014	&	4.299	&	0.043	&	3.404	&	0.041	&	2.946	&	0.016	\\
2458824.8233	&	--	&	--	&	4.996	&	0.011	&	4.314	&	0.069	&	3.433	&	0.028	&	2.978	&	0.012	\\
2458827.8292	&	--	&	--	&	5.002	&	0.009	&	4.316	&	0.038	&	3.424	&	0.027	&	2.921	&	0.010	\\
2458829.8018	&	--	&	--	&	5.000	&	0.012	&	4.316	&	0.035	&	3.411	&	0.025	&	2.962	&	0.019	\\
2458831.8211	&	--	&	--	&	4.994	&	0.018	&	4.310	&	0.059	&	3.394	&	0.040	&	2.956	&	0.026	\\
2458834.8364	&	--	&	--	&	4.981	&	0.006	&	4.293	&	0.028	&	3.407	&	0.023	&	2.914	&	0.014	\\
2458837.7721	&	--	&	--	&	4.984	&	0.013	&	4.300	&	0.039	&	3.401	&	0.041	&	2.912	&	0.026	\\
2458838.8118	&	--	&	--	&	4.987	&	0.006	&	4.299	&	0.042	&	3.408	&	0.032	&	2.926	&	0.012	\\
2458840.8472	&	--	&	--	&	4.995	&	0.013	&	4.312	&	0.055	&	3.432	&	0.034	&	2.956	&	0.018	\\
2458841.7100	&	--	&	--	&	4.992	&	0.022	&	4.316	&	0.101	&	3.432	&	0.073	&	2.933	&	0.029	\\
2458845.8400	&	--	&	--	&	4.994	&	0.011	&	4.318	&	0.055	&	3.424	&	0.026	&	2.924	&	0.013	\\
2458852.7322	&	--	&	--	&	4.966	&	0.013	&	4.287	&	0.032	&	3.404	&	0.032	&	2.853	&	0.011	\\
2458852.8698	&	--	&	--	&	4.963	&	0.008	&	4.271	&	0.042	&	3.401	&	0.030	&	2.921	&	0.020	\\
2458853.8257	&	--	&	--	&	4.961	&	0.013	&	4.271	&	0.074	&	3.398	&	0.052	&	2.880	&	0.019	\\
2458855.8409	&	--	&	--	&	4.960	&	0.012	&	4.269	&	0.033	&	3.392	&	0.053	&	2.869	&	0.017	\\
2458864.6027	&	--	&	--	&	4.939	&	0.014	&	4.246	&	0.047	&	3.363	&	0.036	&	2.872	&	0.024	\\
2458866.6454	&	--	&	--	&	4.932	&	0.045	&	4.236	&	0.042	&	3.383	&	0.031	&	2.770	&	0.032	\\
2458869.8337	&	--	&	--	&	4.910	&	0.011	&	4.228	&	0.064	&	3.352	&	0.038	&	2.865	&	0.019	\\
2458871.5989	&	--	&	--	&	4.886	&	0.016	&	4.203	&	0.072	&	3.327	&	0.040	&	2.839	&	0.012	\\
2458873.6334	&	--	&	--	&	4.880	&	0.029	&	4.170	&	0.031	&	3.349	&	0.022	&	2.797	&	0.033	\\
2458876.7000	&	--	&	--	&	4.873	&	0.023	&	4.175	&	0.053	&	3.330	&	0.030	&	2.828	&	0.019	\\
2458878.6691	&	--	&	--	&	4.868	&	0.013	&	4.178	&	0.061	&	3.359	&	0.031	&	2.807	&	0.025	\\
2458879.7307	&	--	&	--	&	4.845	&	0.012	&	4.150	&	0.086	&	3.299	&	0.030	&	2.790	&	0.020	\\
2458880.6964	&	--	&	--	&	4.843	&	0.016	&	4.096	&	0.168	&	3.351	&	0.029	&	2.789	&	0.026	\\
2458883.6371	&	--	&	--	&	4.785	&	0.009	&	4.097	&	0.068	&	3.273	&	0.014	&	2.749	&	0.021	\\
2458884.6562	&	--	&	--	&	4.776	&	0.017	&	4.108	&	0.054	&	3.321	&	0.023	&	2.736	&	0.018	\\
2458885.6543	&	--	&	--	&	4.762	&	0.017	&	4.095	&	0.056	&	3.275	&	0.067	&	2.713	&	0.026	\\
2458886.6785	&	--	&	--	&	4.769	&	0.014	&	4.081	&	0.060	&	3.280	&	0.015	&	2.733	&	0.028	\\
2458893.6713	&	--	&	--	&	4.856	&	0.010	&	4.176	&	0.041	&	3.374	&	0.032	&	2.842	&	0.019	\\
2458894.7231	&	--	&	--	&	4.885	&	0.012	&	4.187	&	0.050	&	3.376	&	0.030	&	2.829	&	0.009	\\
2458895.6458	&	--	&	--	&	4.907	&	0.009	&	4.220	&	0.063	&	3.386	&	0.036	&	2.846	&	0.019	\\
2458897.6451	&	--	&	--	&	4.954	&	0.012	&	4.265	&	0.050	&	3.453	&	0.043	&	2.894	&	0.017	\\
2458898.6548	&	--	&	--	&	4.967	&	0.009	&	4.279	&	0.012	&	3.451	&	0.018	&	2.896	&	0.016	\\
2458899.6924	&	--	&	--	&	4.981	&	0.007	&	4.287	&	0.033	&	3.464	&	0.025	&	2.954	&	0.019	\\
2458900.6370	&	--	&	--	&	4.971	&	0.006	&	4.286	&	0.036	&	3.446	&	0.022	&	2.930	&	0.007	\\
2458901.5904	&	--	&	--	&	4.977	&	0.008	&	4.289	&	0.029	&	3.480	&	0.024	&	2.934	&	0.014	\\
2458902.7164	&	--	&	--	&	4.988	&	0.010	&	4.297	&	0.057	&	3.480	&	0.019	&	2.950	&	0.013	\\
2458903.5552	&	--	&	--	&	4.987	&	0.012	&	4.295	&	0.062	&	3.459	&	0.051	&	2.922	&	0.010	\\
2458904.5717	&	--	&	--	&	4.992	&	0.013	&	4.290	&	0.042	&	3.487	&	0.041	&	2.948	&	0.026	\\
2458905.5938	&	--	&	--	&	4.985	&	0.011	&	4.291	&	0.035	&	3.469	&	0.021	&	2.935	&	0.009	\\
2458906.5856	&	--	&	--	&	4.992	&	0.011	&	4.302	&	0.023	&	3.473	&	0.019	&	2.941	&	0.014	\\
2458913.5973	&	--	&	--	&	4.987	&	0.013	&	4.294	&	0.042	&	3.456	&	0.049	&	2.958	&	0.012	\\
2458914.5165	&	--	&	--	&	4.986	&	0.018	&	4.302	&	0.046	&	3.484	&	0.038	&	2.948	&	0.014	\\
2458915.5155	&	--	&	--	&	4.999	&	0.010	&	4.316	&	0.026	&	3.469	&	0.029	&	2.974	&	0.013	\\
2458918.5155	&	--	&	--	&	4.983	&	0.007	&	4.296	&	0.037	&	3.482	&	0.017	&	2.940	&	0.008	\\
2458920.5042	&	--	&	--	&	4.998	&	0.020	&	4.296	&	0.050	&	3.467	&	0.031	&	2.929	&	0.020	\\
2458921.8253	&	--	&	--	&	4.985	&	0.013	&	4.282	&	0.049	&	3.466	&	0.038	&	2.933	&	0.028	\\
2458922.6636	&	--	&	--	&	4.982	&	0.009	&	4.292	&	0.015	&	3.446	&	0.037	&	2.979	&	0.021	\\
2458923.5137	&	--	&	--	&	4.963	&	0.011	&	4.282	&	0.035	&	3.441	&	0.019	&	2.937	&	0.016	\\
2458924.5148	&	--	&	--	&	4.977	&	0.011	&	4.280	&	0.036	&	3.466	&	0.035	&	2.932	&	0.019	\\
2458925.4973	&	--	&	--	&	4.958	&	0.020	&	4.290	&	0.061	&	3.482	&	0.078	&	2.898	&	0.033	\\
2458927.4970	&	--	&	--	&	4.976	&	0.033	&	4.263	&	0.046	&	3.415	&	0.034	&	2.943	&	0.036	\\
2458928.4964	&	--	&	--	&	4.957	&	0.009	&	4.264	&	0.032	&	3.457	&	0.026	&	2.904	&	0.022	\\
2459268.5787	&	--	&	--	&	4.731	&	0.019	&	4.079	&	0.045	&	3.265	&	0.023	&	2.791	&	0.014	\\
2459270.5309	&	--	&	--	&	4.721	&	0.010	&	4.083	&	0.053	&	3.245	&	0.038	&	2.790	&	0.015	\\
2459272.6365	&	--	&	--	&	4.736	&	0.011	&	4.088	&	0.013	&	3.226	&	0.031	&	2.831	&	0.018	\\
2459284.5163	&	--	&	--	&	4.736	&	0.013	&	4.096	&	0.047	&	3.270	&	0.032	&	2.834	&	0.014	\\
2459289.5082	&	--	&	--	&	4.723	&	0.008	&	4.085	&	0.034	&	3.263	&	0.026	&	2.784	&	0.016	\\
2459296.4894	&	--	&	--	&	4.745	&	0.006	&	4.104	&	0.051	&	3.241	&	0.032	&	2.839	&	0.019	\\
2459300.4857	&	--	&	--	&	4.762	&	0.013	&	4.116	&	0.034	&	3.267	&	0.021	&	2.843	&	0.017	\\
2459302.4785	&	--	&	--	&	--	&	--	&	--	&	--	&	3.230	&	0.046	&	2.848	&	0.018	\\
2459304.4819	&	--	&	--	&	4.753	&	0.005	&	4.117	&	0.072	&	3.252	&	0.046	&	2.837	&	0.019	\\
2459306.4767	&	--	&	--	&	4.762	&	0.010	&	4.118	&	0.022	&	3.274	&	0.016	&	2.867	&	0.010	\\
2459308.4720	&	--	&	--	&	4.756	&	0.015	&	4.103	&	0.023	&	3.228	&	0.054	&	--	&	--	\\
2459311.4756	&	--	&	--	&	4.774	&	0.017	&	--	&	--	&	--	&	--	&	--	&	--	\\
2459315.4827	&	--	&	--	&	4.747	&	0.017	&	4.112	&	0.077	&	3.250	&	0.053	&	2.817	&	0.029	\\
2459319.4803	&	--	&	--	&	4.743	&	0.037	&	4.111	&	0.096	&	--	&	--	&	2.766	&	0.044	\\
2459366.9450	&	--	&	--	&	4.685	&	0.005	&	4.046	&	0.120	&	3.245	&	0.056	&	2.856	&	0.021	\\
2459394.4550	&	--	&	--	&	4.713	&	0.009	&	--	&		&	--	&	--	&	--	&	--	\\
2459399.4555	&	--	&	--	&	4.696	&	0.012	&	4.057	&	0.062	&	3.231	&	0.042	&	2.773	&	0.032	\\
2459401.4482	&	--	&	--	&	--	&	--	&	4.043	&	0.051	&	3.230	&	0.060	&	2.826	&	0.026	\\
2459402.9596	&	--	&	--	&	4.691	&	0.033	&	4.057	&	0.053	&	3.237	&	0.037	&	2.789	&	0.017	\\
2459404.4558	&	--	&	--	&	4.701	&	0.010	&	4.048	&	0.022	&	3.228	&	0.028	&	2.784	&	0.024	\\
2459407.4507	&	--	&	--	&	4.698	&	0.009	&	4.052	&	0.079	&	3.240	&	0.038	&	2.793	&	0.011	\\
2459412.4510	&	--	&	--	&	4.705	&	0.032	&	4.059	&	0.059	&	3.204	&	0.066	&	2.815	&	0.035	\\
2459416.4677	&	--	&	--	&	4.701	&	0.018	&	4.054	&	0.038	&	3.257	&	0.026	&	2.809	&	0.042	\\
2459422.7115	&	--	&	--	&	4.689	&	0.013	&	4.068	&	0.038	&	3.222	&	0.050	&	2.784	&	0.032	\\
2459436.2116	&	--	&	--	&	4.688	&	0.014	&	4.044	&	0.046	&	3.209	&	0.044	&	2.804	&	0.011	\\
2459437.4560	&	--	&	--	&	4.696	&	0.019	&	4.063	&	0.059	&	3.232	&	0.042	&	2.803	&	0.032	\\
2459526.8080	&	--	&	--	&	4.610	&	0.014	&	3.980	&	0.053	&	3.198	&	0.037	&	2.773	&	0.019	\\
2459530.8073	&	--	&	--	&	4.604	&	0.022	&	3.968	&	0.093	&	3.175	&	0.058	&	2.756	&	0.043	\\
2459534.7970	&	--	&	--	&	4.762	&	0.013	&	4.116	&	0.034	&	3.267	&	0.021	&	2.843	&	0.017	\\
2459537.7962	&	--	&	--	&	4.762	&	0.013	&	4.116	&	0.034	&	3.267	&	0.021	&	2.843	&	0.017	\\
2459542.7795	&	--	&	--	&	4.640	&	0.018	&	4.001	&	0.057	&	3.183	&	0.063	&	2.743	&	0.027	\\
2459544.7745	&	--	&	--	&	4.655	&	0.038	&	4.033	&	0.055	&	3.203	&	0.051	&	2.761	&	0.035	\\
2459550.7993	&	--	&	--	&	4.649	&	0.019	&	4.004	&	0.043	&	3.220	&	0.030	&	2.764	&	0.029	\\
2459551.8175	&	--	&	--	&	4.654	&	0.013	&	3.999	&	0.032	&	3.185	&	0.059	&	2.798	&	0.017	\\
2459567.7039	&	--	&	--	&	4.607	&	0.027	&	3.977	&	0.046	&	3.138	&	0.041	&	2.745	&	0.026	\\
2459569.7287	&	--	&	--	&	4.605	&	0.023	&	3.985	&	0.041	&	3.157	&	0.031	&	2.735	&	0.035	\\
2459597.5868	&	--	&	--	&	4.635	&	0.031	&	3.993	&	0.090	&	3.198	&	0.094	&	2.742	&	0.041	\\
2459599.5861	&	--	&	--	&	4.624	&	0.019	&	3.984	&	0.052	&	3.146	&	0.049	&	2.745	&	0.043	\\
2459698.4770	&	--	&	--	&	4.610	&	0.014	&	3.964	&	0.028	&	3.145	&	0.042	&	2.737	&	0.024	\\
2459701.4820	&	--	&	--	&	4.613	&	0.014	&	3.970	&	0.020	&	3.149	&	0.021	&	2.704	&	0.027	\\
2459705.6290	&	--	&	--	&	4.625	&	0.013	&	3.973	&	0.058	&	3.155	&	0.036	&	2.723	&	0.026	\\
2459709.5410	&	--	&	--	&	4.648	&	0.020	&	4.018	&	0.050	&	3.187	&	0.045	&	2.750	&	0.031	\\
2459712.4739	&	--	&	--	&	4.639	&	0.011	&	4.007	&	0.070	&	3.206	&	0.034	&	2.783	&	0.015	\\
2459716.5575	&	--	&	--	&	4.621	&	0.019	&	3.995	&	0.112	&	3.195	&	0.054	&	2.820	&	0.033	\\
2459719.4731	&	--	&	--	&	4.640	&	0.007	&	4.003	&	0.061	&	3.191	&	0.031	&	2.790	&	0.015	\\
2459723.5523	&	--	&	--	&	4.650	&	0.009	&	4.031	&	0.035	&	3.201	&	0.022	&	2.766	&	0.009	\\
2459726.4629	&	--	&	--	&	4.660	&	0.010	&	4.017	&	0.021	&	3.203	&	0.016	&	2.792	&	0.012	\\
2459740.4623	&	--	&	--	&	4.653	&	0.018	&	4.007	&	0.082	&	3.221	&	0.049	&	2.843	&	0.014	\\
2459747.4755	&	--	&	--	&	4.630	&	0.004	&	3.986	&	0.040	&	3.198	&	0.024	&	2.766	&	0.014	\\
2459751.5386	&	--	&	--	&	4.638	&	0.031	&	3.990	&	0.052	&	3.190	&	0.028	&	2.786	&	0.019	\\
2459761.4600	&	--	&	--	&	4.605	&	0.007	&	3.971	&	0.012	&	3.164	&	0.011	&	2.766	&	0.006	\\
2459768.4629	&	--	&	--	&	4.579	&	0.015	&	3.948	&	0.024	&	3.170	&	0.025	&	2.741	&	0.011	\\
2459772.5189	&	--	&	--	&	4.575	&	0.015	&	3.925	&	0.037	&	3.144	&	0.024	&	2.703	&	0.019	\\
2459779.4619	&	--	&	--	&	4.597	&	0.013	&	3.962	&	0.045	&	3.175	&	0.044	&	2.745	&	0.025	\\
2459782.4708	&	--	&	--	&	4.600	&	0.009	&	3.967	&	0.026	&	3.170	&	0.037	&	2.751	&	0.011	\\
2459786.4634	&	--	&	--	&	4.594	&	0.012	&	3.970	&	0.035	&	3.155	&	0.041	&	2.740	&	0.015	\\
2459790.4628	&	--	&	--	&	4.608	&	0.008	&	3.973	&	0.038	&	3.164	&	0.020	&	2.757	&	0.018	\\
2459810.4769	&	--	&	--	&	4.653	&	0.029	&	4.028	&	0.065	&	3.259	&	0.071	&	2.864	&	0.028	\\
2459811.4635	&	--	&	--	&	4.646	&	0.016	&	4.028	&	0.061	&	3.262	&	0.065	&	2.835	&	0.012	\\
2459812.4578	&	--	&	--	&	4.646	&	0.030	&	4.026	&	0.055	&	3.219	&	0.036	&	2.852	&	0.044	\\
2459813.4593	&	--	&	--	&	4.631	&	0.028	&	4.019	&	0.048	&	3.265	&	0.067	&	2.875	&	0.035	\\
2459814.4631	&	--	&	--	&	4.636	&	0.018	&	4.034	&	0.060	&	3.227	&	0.060	&	2.846	&	0.029	\\
2459872.8551	&	--	&	--	&	4.591	&	0.008	&	3.970	&	0.030	&	3.144	&	0.030	&	2.727	&	0.039	\\
2459875.8345	&	--	&	--	&	4.590	&	0.027	&	3.944	&	0.059	&	3.154	&	0.057	&	2.750	&	0.042	\\
2459877.8332	&	--	&	--	&	4.599	&	0.023	&	3.970	&	0.059	&	3.171	&	0.072	&	2.731	&	0.021	\\
2459880.8161	&	--	&	--	&	4.615	&	0.013	&	3.957	&	0.042	&	3.157	&	0.067	&	2.727	&	0.017	\\
2459884.8428	&	--	&	--	&	4.592	&	0.014	&	3.956	&	0.062	&	3.172	&	0.039	&	2.741	&	0.029	\\
2459887.8235	&	--	&	--	&	4.597	&	0.015	&	3.958	&	0.036	&	3.152	&	0.023	&	2.747	&	0.031	\\
2459894.8250	&	--	&	--	&	4.618	&	0.019	&	3.971	&	0.038	&	3.149	&	0.050	&	2.738	&	0.014	\\
2459897.7907	&	--	&	--	&	4.625	&	0.019	&	3.997	&	0.118	&	3.201	&	0.078	&	2.766	&	0.037	\\
2459901.7767	&	--	&	--	&	4.624	&	0.028	&	3.982	&	0.053	&	3.167	&	0.031	&	2.762	&	0.028	\\
2459905.7739	&	--	&	--	&	4.617	&	0.013	&	3.981	&	0.022	&	3.183	&	0.023	&	2.774	&	0.024	\\
2459912.7568	&	--	&	--	&	4.588	&	0.021	&	3.970	&	0.054	&	3.131	&	0.026	&	2.762	&	0.015	\\
2459919.7534	&	--	&	--	&	4.629	&	0.015	&	3.989	&	0.041	&	3.185	&	0.019	&	2.774	&	0.048	\\
2459926.7445	&	--	&	--	&	4.622	&	0.011	&	4.005	&	0.031	&	3.195	&	0.015	&	2.785	&	0.018	\\
2459929.7120	&	--	&	--	&	4.602	&	0.015	&	3.980	&	0.027	&	3.220	&	0.018	&	2.764	&	0.012	\\
2459933.7428	&	--	&	--	&	4.598	&	0.016	&	3.950	&	0.059	&	--	&	--	&	2.781	&	0.045	\\
2459937.8056	&	--	&	--	&	4.562	&	0.012	&	3.945	&	0.028	&	3.145	&	0.021	&	2.708	&	0.018	\\
2459940.7233	&	--	&	--	&	4.550	&	0.010	&	3.915	&	0.040	&	3.131	&	0.023	&	2.726	&	0.011	\\
2459971.6101	&	--	&	--	&	4.557	&	0.019	&	3.912	&	0.044	&	3.121	&	0.037	&	2.658	&	0.020	\\
2459975.5833	&	--	&	--	&	4.566	&	0.012	&	3.933	&	0.058	&	3.116	&	0.053	&	2.743	&	0.028	\\
2459978.5996	&	--	&	--	&	4.579	&	0.014	&	3.958	&	0.058	&	3.147	&	0.031	&	2.740	&	0.032	\\
2459982.6062	&	--	&	--	&	4.603	&	0.014	&	3.953	&	0.039	&	3.129	&	0.038	&	2.761	&	0.020	\\
2459989.5723	&	--	&	--	&	4.600	&	0.019	&	3.955	&	0.076	&	3.134	&	0.069	&	2.692	&	0.046	\\
2459992.6121	&	--	&	--	&	4.597	&	0.019	&	3.976	&	0.074	&	3.168	&	0.044	&	2.719	&	0.018	\\
2460003.5210	&	--	&	--	&	4.604	&	0.021	&	3.963	&	0.069	&	3.167	&	0.038	&	2.784	&	0.040	\\
2460010.5114	&	--	&	--	&	4.582	&	0.025	&	3.968	&	0.036	&	3.167	&	0.031	&	2.738	&	0.023	\\
2460013.5159	&	--	&	--	&	4.579	&	0.024	&	3.955	&	0.044	&	3.142	&	0.048	&	2.721	&	0.025	\\
2460017.6010	&	--	&	--	&	4.566	&	0.026	&	3.939	&	--	&	3.130	&	0.025	&	--	&	0.017	\\
2460020.5064	&	--	&	--	&	4.566	&	0.021	&	3.944	&	0.069	&	3.177	&	0.055	&	2.775	&	0.035	\\
2460023.5043	&	--	&	--	&	4.569	&	0.023	&	3.952	&	0.063	&	3.157	&	0.034	&	2.687	&	0.034	\\
2460031.5199	&	--	&	--	&	4.558	&	0.031	&	3.935	&	0.048	&	3.133	&	0.029	&	2.763	&	0.035	\\
2460036.4908	&	--	&	--	&	4.558	&	0.030	&	3.950	&	0.044	&	3.145	&	0.021	&	2.697	&	0.037	\\
2460038.5232	&	--	&	--	&	4.565	&	0.007	&	3.937	&	0.039	&	3.138	&	0.021	&	2.709	&	0.018	\\
2460041.5526	&	--	&	--	&	4.589	&	0.015	&	3.958	&	--	&	3.154	&	0.039	&	--	&	0.018	\\
2460043.4840	&	--	&	--	&	4.586	&	0.032	&	3.974	&	0.037	&	3.131	&	0.046	&	2.738	&	0.060	\\
2460051.5404	&	--	&	--	&	4.557	&	0.017	&	3.941	&	0.043	&	3.141	&	0.073	&	2.744	&	0.021	\\
2460054.4788	&	--	&	--	&	4.554	&	0.020	&	3.934	&	0.034	&	3.138	&	0.021	&	2.717	&	0.046	\\
2460059.4777	&	--	&	--	&	4.554	&	0.014	&	3.913	&	0.020	&	3.133	&	0.017	&	2.697	&	0.015	\\
2460062.4718	&	--	&	--	&	4.564	&	0.012	&	3.950	&	0.045	&	3.139	&	0.053	&	2.724	&	0.030	\\
2460066.4726	&	--	&	--	&	4.565	&	0.009	&	3.922	&	0.050	&	3.142	&	0.037	&	2.705	&	0.015	\\
2460069.4727	&	--	&	--	&	4.558	&	0.017	&	3.927	&	0.039	&	3.135	&	0.021	&	2.703	&	0.049	\\
2460073.4618	&	--	&	--	&	4.565	&	0.014	&	3.916	&	0.047	&	3.131	&	0.033	&	2.734	&	0.019	\\
2460076.4664	&	--	&	--	&	4.574	&	0.009	&	3.927	&	0.069	&	3.114	&	0.041	&	2.696	&	0.018	\\
2460080.5042	&	--	&	--	&	4.587	&	0.012	&	3.937	&	--	&	--	&	--	&	--	&	 -	\\
2460085.4677	&	--	&	--	&	4.560	&	0.012	&	3.932	&	0.040	&	3.124	&	0.033	&	2.720	&	0.025	\\
2460089.4840	&	--	&	--	&	4.564	&	0.017	&	3.917	&	0.032	&	3.145	&	0.043	&	2.680	&	0.019	\\
2460094.4543	&	--	&	--	&	4.588	&	0.028	&	3.956	&	0.069	&	3.150	&	0.056	&	2.783	&	0.056	\\
2460097.5110	&	--	&	--	&	4.587	&	0.014	&	3.953	&	0.041	&	3.130	&	0.019	&	2.689	&	0.014	\\
2460101.4549	&	--	&	--	&	4.594	&	0.017	&	3.941	&	0.071	&	3.102	&	0.047	&	2.733	&	0.024	\\
2460111.4751	&	--	&	--	&	4.614	&	0.023	&	3.973	&	0.044	&	3.152	&	0.034	&	2.721	&	0.054	\\
2460170.4588	&	--	&	--	&	4.579	&	0.024	&	3.953	&	0.098	&	3.142	&	0.098	&	2.715	&	0.055	\\
2460233.8704	&	--	&	--	&	4.660	&	0.016	&	4.033	&	0.045	&	3.193	&	0.036	&	2.824	&	0.032	\\
2460236.8482	&	--	&	--	&	4.662	&	0.017	&	4.037	&	0.042	&	3.197	&	0.043	&	2.789	&	0.017	\\
2460243.8351	&	--	&	--	&	4.656	&	0.014	&	4.049	&	0.043	&	3.204	&	0.017	&	2.831	&	0.019	\\
2460250.8154	&	--	&	--	&	4.635	&	0.023	&	4.011	&	0.059	&	3.196	&	0.037	&	2.790	&	0.018	\\
2460264.8462	&	--	&	--	&	4.649	&	0.013	&	4.012	&	0.050	&	3.206	&	0.045	&	2.783	&	0.018	\\
2460271.8058	&	--	&	--	&	4.641	&	0.014	&	3.999	&	0.058	&	3.189	&	0.072	&	2.778	&	0.023	\\
2460285.8088	&	--	&	--	&	4.647	&	0.013	&	4.023	&	0.057	&	3.210	&	0.038	&	2.762	&	0.011	\\
2460292.7960	&	--	&	--	&	4.606	&	0.008	&	3.979	&	0.029	&	3.162	&	0.043	&	2.743	&	0.017	\\
2460299.8014	&	--	&	--	&	4.601	&	0.009	&	3.965	&	0.034	&	3.152	&	0.036	&	2.753	&	0.016	\\
2460313.7251	&	--	&	--	&	4.627	&	0.009	&	4.016	&	0.048	&	3.182	&	0.032	&	2.755	&	0.011	\\
2460327.6945	&	--	&	--	&	4.646	&	0.023	&	4.022	&	0.055	&	3.182	&	0.054	&	2.794	&	0.025	\\
2460334.6047	&	--	&	--	&	4.649	&	0.017	&	4.012	&	0.044	&	3.190	&	0.049	&	2.786	&	0.024	\\
2460342.7481	&	--	&	--	&	4.649	&	0.010	&	4.021	&	0.056	&	3.196	&	0.035	&	2.782	&	0.014	\\
2460346.6060	&	--	&	--	&	4.649	&	0.026	&	4.019	&	0.062	&	3.206	&	0.066	&	2.820	&	0.034	\\
2460350.5536	&	--	&	--	&	4.640	&	0.015	&	4.019	&	0.032	&	3.209	&	0.046	&	2.792	&	0.018	\\
2460354.5294	&	--	&	--	&	4.638	&	0.014	&	4.005	&	0.043	&	3.203	&	0.036	&	2.806	&	0.028	\\
2460356.5567	&	--	&	--	&	4.627	&	0.009	&	3.995	&	0.049	&	3.187	&	0.034	&	2.753	&	0.029	\\
2460360.5411	&	--	&	--	&	4.627	&	0.007	&	3.993	&	0.028	&	3.160	&	0.023	&	2.751	&	0.011	\\
2460367.5242	&	--	&	--	&	4.617	&	0.013	&	3.970	&	0.050	&	3.130	&	0.041	&	2.753	&	0.040	\\
2460370.5160	&	--	&	--	&	4.629	&	0.013	&	3.990	&	0.037	&	3.154	&	0.050	&	2.738	&	0.026	\\
2460377.5186	&	--	&	--	&	4.647	&	0.029	&	4.004	&	0.054	&	3.209	&	0.050	&	2.756	&	0.023	\\
2460384.5052	&	--	&	--	&	4.661	&	0.025	&	4.011	&	0.065	&	3.152	&	0.064	&	2.764	&	0.024	\\
2460391.5119	&	--	&	--	&	4.657	&	0.015	&	4.024	&	0.063	&	3.179	&	0.055	&	2.793	&	0.030	\\
2460395.5096	&	--	&	--	&	4.642	&	0.014	&	4.004	&	0.046	&	3.181	&	0.039	&	2.745	&	0.021	\\
2460398.5785	&	--	&	--	&	4.624	&	0.013	&	3.991	&	0.099	&	3.191	&	0.053	&	2.773	&	0.017	\\
2460403.2411	&	--	&	--	&	4.617	&	0.026	&	3.993	&	0.034	&	3.146	&	0.047	&	2.742	&	0.026	\\
2460405.4923	&	--	&	--	&	4.619	&	0.011	&	3.981	&	0.033	&	3.161	&	0.044	&	--	&	--	\\
2460409.4849	&	--	&	--	&	4.615	&	0.013	&	3.958	&	0.039	&	3.146	&	0.028	&	2.714	&	0.019	\\
2460412.5158	&	--	&	--	&	4.570	&	0.012	&	3.943	&	0.035	&	3.143	&	0.047	&	2.732	&	0.026	\\
2460416.5072	&	--	&	--	&	4.576	&	0.011	&	3.937	&	0.021	&	3.139	&	0.028	&	2.690	&	0.019	\\
2460419.4822	&	--	&	--	&	4.593	&	0.015	&	3.941	&	0.053	&	3.138	&	0.049	&	2.710	&	0.028	\\
2460423.5228	&	--	&	--	&	4.622	&	0.014	&	3.959	&	0.045	&	3.147	&	0.048	&	--	&	--	\\
2460430.4784	&	--	&	--	&	4.612	&	0.005	&	3.962	&	0.037	&	3.159	&	0.028	&	2.724	&	0.016	\\
2460433.4687	&	--	&	--	&	4.607	&	0.010	&	3.960	&	0.053	&	3.151	&	0.019	&	2.686	&	0.015	\\
2460437.5251	&	--	&	--	&	4.628	&	0.007	&	3.975	&	0.028	&	3.153	&	0.024	&	2.743	&	0.011	\\
2460440.4637	&	--	&	--	&	4.641	&	0.010	&	3.999	&	0.037	&	3.173	&	0.031	&	2.746	&	0.019	\\
2460444.4585	&	--	&	--	&	4.648	&	0.011	&	3.998	&	0.041	&	3.187	&	0.018	&	2.748	&	0.007	\\
2460447.4886	&	--	&	--	&	4.638	&	0.007	&	4.002	&	0.028	&	3.167	&	0.041	&	2.752	&	0.009	\\
2460451.4694	&	--	&	--	&	4.642	&	0.010	&	4.005	&	0.025	&	3.174	&	0.017	&	2.759	&	0.010	\\
2460454.4677	&	--	&	--	&	4.639	&	0.004	&	3.990	&	0.029	&	3.181	&	0.020	&	2.749	&	0.019	\\
2460458.4754	&	--	&	--	&	4.657	&	0.008	&	4.015	&	0.034	&	3.169	&	0.014	&	2.745	&	0.019	\\
2460461.4707	&	--	&	--	&	4.647	&	0.014	&	3.995	&	0.046	&	3.181	&	0.018	&	2.760	&	0.023	\\
2460464.4776	&	--	&	--	&	4.659	&	0.019	&	3.996	&	0.033	&	3.158	&	0.041	&	2.786	&	0.010	\\
2460467.4722	&	--	&	--	&	4.633	&	0.009	&	3.994	&	0.026	&	3.157	&	0.027	&	2.744	&	0.007	\\
2460479.5583	&	--	&	--	&	4.612	&	0.028	&	3.983	&	0.034	&	3.133	&	0.056	&	2.703	&	0.033	\\
2460482.4917	&	--	&	--	&	4.629	&	0.012	&	3.979	&	0.057	&	3.170	&	0.066	&	2.722	&	0.021	\\
2460488.4523	&	--	&	--	&	4.639	&	0.012	&	3.978	&	0.040	&	3.147	&	0.047	&	2.728	&	0.018	\\
2460510.4589	&	--	&	--	&	4.619	&	0.025	&	3.979	&	0.073	&	3.149	&	0.064	&	2.751	&	0.041	\\
2460530.4571	&	--	&	--	&	4.650	&	0.013	&	3.984	&	0.065	&	3.141	&	0.040	&	--	&	--	\\
2460531.4729	&	--	&	--	&	4.639	&	0.019	&	3.994	&	0.080	&	--	&	--	&	2.716	&	0.040	\\
2460532.4525	&	--	&	--	&	4.656	&	0.027	&	--	&	--	&	3.188	&	0.038	&	2.794	&	0.030	\\
2460608.8450	&	--	&	--	&	4.666	&	0.013	&	4.023	&	0.022	&	3.188	&	0.078	&	2.773	&	0.019	\\
2460612.8194	&	4.446	&	0.008	&	4.679	&	0.007	&	4.446	&	0.027	&	3.156	&	0.016	&	2.723	&	0.008	\\
2460630.7851	&	4.426	&	0.003	&	4.651	&	0.025	&	4.426	&	0.036	&	3.195	&	0.051	&	2.747	&	0.003	\\
2460641.7431	&	4.428	&	0.007	&	4.644	&	0.032	&	4.428	&	0.108	&	3.143	&	0.072	&	2.694	&	0.007	\\
2460671.8230	&	4.465	&	0.003	&	4.692	&	0.017	&	4.465	&	0.085	&	3.193	&	0.025	&	2.776	&	0.003	\\
2460675.6720	&	4.454	&	0.003	&	4.680	&	0.015	&	4.454	&	0.041	&	3.173	&	0.057	&	2.770	&	0.003	\\
2460682.8000	&	4.483	&	0.004	&	4.710	&	0.021	&	4.483	&	0.024	&	3.193	&	0.016	&	2.790	&	0.004	\\
2460685.6726	&	4.467	&	0.010	&	4.700	&	0.034	&	4.467	&	0.079	&	3.203	&	0.054	&	2.671	&	0.010	\\
2460699.6137	&	4.481	&	0.007	&	4.702	&	0.015	&	4.481	&	0.050	&	3.193	&	0.033	&	2.758	&	0.007	\\
2460703.6273	&	--	&	--	&	4.700	&	0.012	&	--	&	--	&	3.167	&	0.055	&	--	&	--	\\
2460706.5777	&	--	&	--	&	4.696	&	0.010	&	--	&	--	&	3.174	&	0.033	&	2.759	&	0.008	\\
2460710.5683	&	4.467	&	0.007	&	4.691	&	0.025	&	4.467	&	0.056	&	3.142	&	0.049	&	2.727	&	0.007	\\
2460712.5493	&	4.470	&	0.006	&	4.685	&	0.020	&	4.470	&	0.049	&	3.181	&	0.048	&	2.745	&	0.006	\\
2460849.4569	&	--	&	--	&	4.632	&	0.013	&	--	&	0.012	&	3.115	&	0.017	&	2.653	&	0.007	\\
2460850.4412	&	--	&	--	&	4.631	&	0.011	&	--	&	--	&	--	&	--	&	--	&	--	\\
2460851.4564	&	4.420	&	0.003	&	4.642	&	0.014	&	4.420	&	0.039	&	3.102	&	0.045	&	2.635	&	0.003	\\
2460853.4517	&	4.412	&	0.009	&	4.618	&	0.036	&	4.412	&	0.035	&	3.059	&	0.042	&	2.602	&	0.009	\\
2460857.4538	&	--	&	--	&	4.639	&	0.016	&	--	&	--	&	--	&	--	&	--	&	--	\\
2460859.4579	&	4.402	&	0.012	&	4.625	&	0.032	&	4.402	&	0.055	&	3.090	&	0.057	&	2.599	&	0.012	\\
2460861.4575	&	4.425	&	0.007	&	4.636	&	0.011	&	4.425	&	0.007	&	3.120	&	0.025	&	2.626	&	0.007	\\
2460864.4620	&	4.425	&	0.004	&	4.625	&	0.012	&	4.425	&	0.031	&	3.106	&	0.056	&	2.614	&	0.004	\\
2460866.4583	&	4.406	&	0.004	&	4.622	&	0.018	&	4.406	&	0.041	&	3.086	&	0.036	&	2.607	&	0.004	\\
2460868.4569	&	4.391	&	0.012	&	4.608	&	0.030	&	4.391	&	0.074	&	3.072	&	0.053	&	2.621	&	0.012	\\
2460871.4686	&	4.393	&	0.010	&	4.605	&	0.020	&	4.393	&	0.063	&	3.077	&	0.049	&	2.621	&	0.010	\\
2460873.4641	&	4.388	&	0.008	&	4.598	&	0.012	&	4.388	&	0.027	&	3.097	&	0.019	&	2.636	&	0.008	\\
2460875.4601	&	4.398	&	0.008	&	4.610	&	0.014	&	4.398	&	0.065	&	3.101	&	0.051	&	2.652	&	0.008	\\
2460877.4615	&	4.403	&	0.006	&	4.599	&	0.020	&	4.403	&	0.071	&	3.061	&	0.041	&	2.610	&	0.006	\\
2460885.4662	&	4.379	&	0.008	&	4.607	&	0.015	&	4.379	&	0.055	&	3.072	&	0.049	&	2.617	&	0.008	\\
\end{longtable}
\noindent\centering\textbf{Notes:} Eta Car differential photometry data from March 2019 to July 2025\par
TELESCOPE: 0.60 m Helen Sawyer Hogg (HSH)\\
CAMERA: CCD, SBIG STL1001E\\
OBSERVATORY: Complejo Astronomico El Leoncito (CASLEO), San Juan, Argentina\\
OBSERVING and PROCESSING: E. Fernandez-Lajus\\
The magnitudes are in the La Plata system.
%\commentFN{Should we place these notes in the text instead?}								
\end{center}

\label{lastpage}
\end{document}